\documentclass[letterpaper,11pt]{article}
\usepackage{hyperref}

\usepackage{amsmath}
\usepackage{amsfonts}
\usepackage{amsthm}
\usepackage{bbm}
\usepackage[only,llbracket,rrbracket]{stmaryrd}

\usepackage{algorithm}
\usepackage{algpseudocode}

\usepackage[style=alphabetic,natbib=true,maxcitenames=3]{biblatex}
\usepackage[margin=1in]{geometry}

\newtheorem{thm}{Theorem}[section]
\newtheorem{cor}[thm]{Corollary}
\newtheorem{prop}[thm]{Proposition}
\newtheorem{lmm}[thm]{Lemma}
\theoremstyle{definition}
\newtheorem{defi}[thm]{Definition}
\theoremstyle{definition}
\newtheorem{que}{Question}

\newcommand{\N}{\mathbb{N}}
\renewcommand{\H}{\mathcal{H}}
\newcommand{\Heff}{\mathcal{H}^\textsf{eff}}
\renewcommand{\P}{\mathbb{P}}
\newcommand{\E}{\mathbb{E}}
\newcommand{\R}{\mathbb{R}}
\newcommand{\Q}{\mathbb{Q}}
\newcommand{\card}[1]{\left| #1 \right|}
\newcommand{\ind}[1]{\mathbbm{1}_{#1}}

\newcommand{\nashe}{\textsc{Nash} equilibrium}
\newcommand{\nashes}{\textsc{Nash} equilibria}

\title{Playing repeated games with sublinear randomness}
\author{Farid Arthaud\\
{\footnotesize MIT CSAIL}\\
{\footnotesize Cambridge, Massachusetts}\\
{\footnotesize \href{mailto:farto@csail.mit.edu}{\texttt{farto@csail.mit.edu}}}}
\date{}

\begin{document}
\maketitle
\thispagestyle{empty}

\begin{abstract}
	We study the amount of entropy players asymptotically need to play a
	repeated normal-form game in a \nashe{}.
	\citet{DBLP:journals/mst/HubacekNU16} (SAGT~'15,~TCSys~'16) gave
	sufficient conditions on a game for the minimal amount of randomness
	required to be \( O(1) \) or \( \Omega(n) \) for all players, where
	\( n \) is the number of repetitions.
	We provide a complete characterization of games in which there exists
	\nashes{} of the repeated game using \( O(1) \) randomness, closing an
	open question posed by~\citet{DBLP:conf/sigecom/BudinichF11}~(EC~'11)
	and~\citet{DBLP:journals/mst/HubacekNU16}.
	Moreover, we show a 0--1 law for randomness in repeated games, showing
	that any repeated game either has \( O(1) \)-randomness \nashes{}, or all
	of its \nashes{} require \( \Omega(n) \) randomness.
	Our techniques are general and naturally characterize the payoff space of
	sublinear-entropy equilibria, and could be of independent interest to
	the study of players with other bounded capabilities in repeated games.
\end{abstract}
\newpage

\tableofcontents
\thispagestyle{empty}
\newpage

\clearpage
\setcounter{page}{1}

\section{Introduction}
We study the amount of entropy required to play a \nashe{} of a finitely repeated
normal-form game.
The seminal result of \textsc{Nash} in game theory states that any normal-form
game has a \nashe{} if each player can randomize their strategy.
As noted in previous
work~\citep{DBLP:conf/sigecom/BudinichF11,DBLP:journals/mst/HubacekNU16,DBLP:journals/jet/HalpernP15},
the assumption that players can randomize arbitrarily is non-trivial, as true
randomness might be scarce or costly and humans are known to have difficulty
generating truly random sequences.

An immediate \nashe{} of any repeated game is to repeat a \nashe{} of the stage
game at each round, irrespective of the outcomes of previous rounds.
When the chosen \nashe{} of the stage game requires a player to
randomize, this strategy requires that player to use an amount of entropy that
grows linearly with the number of repetitions.
However, not all equilibria of the repeated game are of this form: in some cases,
the repeated game has equilibria in which all players use only constant
randomness.
From here, several natural questions arise,
\begin{que}\label{que:q1} When do there exist \nashes{} of the repeated
game in which all players use sublinear entropy?\end{que}
\begin{que}\label{que:q2} When do all equilibria of the repeated game require
all players to use linear entropy?\end{que}
\begin{que}\label{que:q3} What are the answers to the two previous questions
	when only restricting the entropy used by a subset of the
players?\end{que}
\begin{que}\label{que:q4} How does restricting randomization affect the
payoffs of the \nashes{} of the repeated game?\end{que}

\citet{DBLP:conf/sigecom/BudinichF11} first studied the question of equilibria
with bounded randomness in repeated games in the case of the two-player matching
pennies game.
They find that each player needs \( n \) independent random bits to play any
equilibrium of the repeated game, where \( n \) is the number of repetitions.
This further motivates Questions~\ref{que:q1} and~\ref{que:q2} above: it shows
that matching pennies fits in the case specified by Question~\ref{que:q2} and
shows there are simple games where the answer to Question~\ref{que:q1} is
negative.

\citet{DBLP:journals/mst/HubacekNU16} then extended this study to general
repeated games with any number of players.
They identify a sufficient condition on the stage game for there to exist an
equilibrium of the repeated game requiring \( O(1) \) entropy for all players,
therefore providing a sufficient condition for Question~\ref{que:q1}.
They also identify a sufficient condition under which all \nashes{} of the
repeated game require \( \Omega(n) \) entropy for each player, therefore
providing a sufficient condition for Question~\ref{que:q2}.
This latter condition is satisfied by all two-player zero-sum games where all
\nashes{} require all players to randomize, in particular the matching pennies
game.

These results show that there are at least two regimes for the required entropy
to play a \nashe{} of a repeated game (constant or linear).
Moreover, both~\citet{DBLP:conf/sigecom/BudinichF11}
and~\citet{DBLP:journals/mst/HubacekNU16} posed the open question of whether a
characterization of games satisfying Question~\ref{que:q1} can be found.

\subsection{Our results}
We close an open question posed by~\citet{DBLP:conf/sigecom/BudinichF11}
and~\citet{DBLP:journals/mst/HubacekNU16} by completely characterizing
normal-form games for which there are \nashes{} of the repeated game in which all
players use \( O(1) \) randomness as a function of the number of repetitions
(Theorem~\ref{thm:mostgeneral}).
This closes Question~\ref{que:q1} above.

An important consequence of the techniques we use to prove these results is
that we show that when \( O(1) \) randomness is not sufficient, then in all
equilibria at least one player uses entropy \( \Omega(n) \).
As such, the existence of sublinear-randomness \nashes{} of a repeated game
implies the existence of \( O(1) \)-randomness \nashes{}, meaning that
Question~\ref{que:q1} and its analog for constant-randomness equilibria always
have the same answer.
The only previously known result on sublinear-entropy \nashes{} in general
repeated games assumes that one-way functions do not exist and that players are
computationally bounded~\citep{DBLP:journals/mst/HubacekNU16}.
While precise statements require more definitions, we informally summarize the
stated results so far.
\begin{thm}[Informal version of Theorem~\ref{thm:mostgeneral}]
	For a normal-form game \( G \), there are constant-randomness \nashes{}
	(or equivalently, sublinear-randomness \nashes{}) of its repeated games
	if and only if
	there exists a subset \( S' \) of the pure strategy profiles of \( G \)
	and a convex combination \( \gamma_s \) for \( s \in S' \), such that the
	payoff vector \( \sum_{s \in S'} \gamma_s u(s) \) is at least the
	pure-strategy minmax of each player, and each player that is not best
	responding in every strategy in \( S' \) can be punished by a finite
	number of repetitions of \( G \).
\end{thm}

We also close Question~\ref{que:q1} in the case of repeated games with
observable distributions (Theorem~\ref{thm:observable}), a setting in which the
distribution chosen by each player to randomize their actions is revealed after
reach round (Definition~\ref{defi:observable}).
Regarding Question~\ref{que:q3}, for any subset \( B \) of the players, we
provide a sufficient condition and a necessary condition for there to exist a
\nashe{} of the repeated game in which all players in \( B \) use \( O(1) \)
entropy (Theorem~\ref{thm:subset}).

Another consequence of our results is that they naturally characterize the
set of payoffs to which the average payoff of a sublinear-entropy \nashe{} can
converge to, closing Question~\ref{que:q4}.

\paragraph{Two-player games}
For two-player games, our characterization of Question~\ref{que:q1}
(Theorem~\ref{thm:twoplayer}) only requires testing the feasibility of a
polynomial-size linear program over the stage game \( G \) and testing for the
existence of a \nashe{} of \( G \) satisfying certain linear inequalities.
This means the condition is decidable and is even in \textsf{NP} (this is also
true for the observable distributions case with any number of players mentioned
above).
\begin{thm}[Informal version of Theorem~\ref{thm:twoplayer}]
	For a two-player game \( G \), there are constant-randomness \nashes{}
	(or equivalently, sublinear-randomness \nashes{}) of \( G \)'s repeated
	game if and only if
	\( G \) has a pure \nashe{} or there exists a convex combination \(
	\gamma_s \) over pure strategies \( s \in S \) of \( G \) such that the
	payoff vector \( \sum_s \gamma_s u(s) \) is at least the pure-strategy
	minmax of each player and \( G \) has a \nashe{} that is strictly better
	than the minmax of one of the players.
\end{thm}

Still in the case of two-player games, we also close
Questions~\ref{que:q2},~\ref{que:q3} and~\ref{que:q4}: we provide a complete
characterization of games in which both players each require \( \Omega(n) \)
entropy to play any equilibrium (Theorem~\ref{thm:omegatwo}), a complete
characterization of games in which one player only needs sublinear entropy
(Theorem~\ref{thm:subsettwo}) and a characterization of payoffs that are
achievable asymptotically (Corollary~\ref{cor:payofftwo}).
All of these conditions are in \textsf{NP}.

Note above that \citep{DBLP:conf/sigecom/BudinichF11} deal with random bits
whereas \citep{DBLP:journals/mst/HubacekNU16} bound entropy.
We work with entropy (Definition~\ref{defi:entropy}) as
in~\citep{DBLP:journals/mst/HubacekNU16}.
In Subsection~\ref{sec:computational} we introduce a computational model that
precisely relates the expected random bits used by an algorithm implementing a
player's strategy and the entropy as defined by
\citep{DBLP:journals/mst/HubacekNU16}.
A difference between \citep{DBLP:journals/mst/HubacekNU16} and our results is the
distinction between \emph{effective entropy} and \emph{total entropy} of an
equilibrium (also in Definition~\ref{defi:entropy}), which is explained in
Subsection~\ref{sec:related}.

\paragraph{Roadmap}
The remainder of the article is organized as follows.
The remainder of this section contains an overview of our techniques and further
related work.
Section~\ref{sec:prelim} contains the main definitions and introduces our
computational model under which the entropy characterization relates to exact
play of equilibria using balanced random bits.
We progressively build up towards the most general case by presenting our results
on sublinear-entropy \nashes{} of repeated games in Section~\ref{sec:twoplayer}
for two-player games, Section~\ref{sec:observable} for \( m \)-player games with
observable distributions, and Section~\ref{sec:general} for general \( m
\)-player games.
Finally, Appendix~\ref{app:effective} contains a characterization of games in
which there are \nashes{} in which all players need \( O(1) \) effective entropy
(rather than total entropy, see Definition~\ref{defi:entropy} and
Subsection~\ref{sec:related}).

\subsection{Our techniques: playing with sublinear randomness}
Our starting point is the following sufficient condition, which we then weaken
into a necessary (and sufficient) condition.
\begin{prop}[Informal version of Proposition~\ref{prop:a0suff}]\label{prop:a0suffinf}
	If \( G \) has a convex combination \( {\left( \gamma_s \right)}_{s\in S}
	\) over pure strategies \( s \) of \( G \) such that the payoff vector
	\( \sum_s \gamma_s u(s) \) is at least the pure-strategy minmax of each
	player, and every player either is best responding in every strategy \( s
	\) where \( \gamma_s > 0 \) or has a \nashe{} that has a payoff strictly
	better than their minmax value, then there exists \( O(1) \)-entropy
	\nashes{} of \( G \)'s repeated game.
\end{prop}
Note this proposition generalizes
\citep[Theorem~2]{DBLP:journals/mst/HubacekNU16} in two ways: (i) it does not
require all players to have a strictly individually rational NE and (ii) it does
not require the coefficients \( \gamma_s \) to be rationals.
When \( \gamma_s \in \Q \) for all \( s \) (as assumed
by~\citep{DBLP:journals/mst/HubacekNU16}), a construction similar to the one from
the folk theorem of~\citet{benoit1987nash} is possible.
These constructions fail otherwise, which requires us to construct a different
type of equilibria.

We use two main tools to weaken this statement into a necessary condition.
The first shows that the last assumption in the proposition above (that each
player has a \nashe{} strictly better than their minmax) must be true, not of all
players, but only of at least one player.
\begin{prop}[Informal version of Proposition~\ref{prop:simplenec}]\label{prop:simplenecinf}
	Assume \( G \) does not have a pure \nashe{}, and has \nashes{} of its
	repeated game using \( o(n) \) entropy.
	Then \( G \) has a \nashe{} in which at least one player's payoff is
	strictly greater than their minmax.
\end{prop}
The proof uses the compactness of \nashes{} of \( G \) to find a round where a
player is not best responding, and from there finds a future round that must
reward/punish that player.

Let \( v_i \) be the pure-strategy minmax of player~\( i \), i.e.\ the worst
punishment that can be imposed on player~\( i \) without randomizing (see
Definition~\ref{defi:minmax}).
The second tool to weaken the above Proposition~\ref{prop:a0suffinf} is a lemma
that bounds the long-term punishments that can be imposed on a player for
deviating (see Definition~\ref{defi:punish} for a formal definition), when all
players are bounded in their entropy.
\begin{lmm}[Informal version of Lemma~\ref{lmm:avg}]
	For any small enough \( \varepsilon > 0 \) and any \( n \)-round
	punishment of player~\( i \) in which all players use at most \( O(f(n))
	\) entropy, the average payoff of player~\( i \) is at least,
	\begin{equation}\label{eq:worstpunish}
		v_i - O \left( \frac{f(n)}{\varepsilon n} \right) -
		O(\varepsilon).
	\end{equation}
\end{lmm}
The general idea behind the proof is that rounds in the punishment either use
a low amount of entropy or a high amount of entropy: low-entropy rounds cannot
have payoffs too far off from the pure minmax, whereas there cannot be too many
high-entropy rounds.
This bound allows us to prove the following lemma,
which shows that the average payoff of any sublinear-entropy equilibrium
must be at least the pure-strategy minmax.
Intuitively, if the worst punishment possible approximately imposes the payoff \(
v_i \) on player~\( i \), then their average payoff on-path must be at least \(
v_i \) in the long run, in order to avoid players gaining a linear advantage from
deviating.
\begin{lmm}[Informal version of Lemma~\ref{lmm:necfeasible}]\label{lmm:necfeasibleinf}
	If \( G \) has \nashes{} of its repeated game using \( o(n) \)
	randomness, then there exists a convex combination \( \gamma_s \) over
	pure strategies \( s \) of \( G \) such that the payoff vector \( \sum_s
	\gamma_s u(s) \) is at least the pure minmax of each player.
\end{lmm}
Note that this recovers the first assumption made in the sufficient condition of
Proposition~\ref{prop:a0suffinf} above, by showing it is true as soon as there
are sublinear-entropy \nashes{}.
In the two-player case, we show in Theorem~\ref{thm:twoplayer} that this
condition along with the one in Proposition~\ref{prop:simplenecinf} are together
sufficient: this amounts to weakening the assumptions of
Proposition~\ref{prop:a0suffinf} and building a \( O(1) \)-entropy equilibrium
under these weaker assumptions.

The proofs for repeated games with observable distributions
(Section~\ref{sec:observable}) and for general repeated games
(Section~\ref{sec:general}) rely on similar ideas, but require stronger tools to
identify punishments in the necessary condition.
In particular, Proposition~\ref{prop:simplenecinf} is no longer sufficient on its
own, and the proofs of Theorems~\ref{thm:observable} and~\ref{thm:mostgeneral}
must instead produce strategy profiles where a subset of players are best
responding and a particular player's payoff is strictly better than their minmax.
This results in an ordering of the players in their rewards/punishments: after a
player has been punished, they must not have any other opportunities for
deviation during the punishment of other players.

\subsection{Further related work}\label{sec:related}
There is one difference between the setting in which
\citet{DBLP:journals/mst/HubacekNU16}'s results hold and our setting, namely the
definition used for entropy of an equilibrium.
The sufficient condition for \( O(1) \) randomness equilibria
in~\citep[Theorem~2]{DBLP:journals/mst/HubacekNU16} bounds \emph{effective
entropy} rather than total entropy: instead of measuring the amount of entropy
along any path of the game, they only measure entropy along paths which are
sampled with non-zero probability under equilibrium play (see
Definition~\ref{defi:entropy} for formal definitions).
Effective entropy could model a situation in which linear entropy is costly
yet achievable (players agree that they could hurt each other using linear
entropy so they won't deviate), but it does not model situations where players
are unable to produce linear amounts of entropy.
Our sufficient conditions are therefore stronger in this regard, as bounding
total entropy implies the same bound for effective entropy, and
Proposition~\ref{prop:simplesuff} does so under similar assumptions
to~\citep[Theorem~2]{DBLP:journals/mst/HubacekNU16}.
Our techniques are sufficiently general that our characterizations extend to the
case of effective entropy naturally: Appendix~\ref{app:effective} contains
an analog of our most general result for effective entropy.

\paragraph{Costly randomness}
\citet{DBLP:journals/jet/HalpernP15} consider adding a cost of computation to
games, in a setting where players are modeled by \textsc{Turing} machines.
They show that \nashes{} do not necessarily exist when players pay a cost for
randomness, but that whenever randomness is free (but computation can be costly)
\nashes{} always exist.

\paragraph{Off-equilibrium play with bounded randomness}
There is a line of research on the maxmin payoff in repeated games where \(
\Omega(n) \) entropy is needed but only sublinear entropy is available to one
player, and specifically in two-player zero-sum games.
\citet{DBLP:conf/sigecom/BudinichF11} had already shown that in repeated matching
pennies, if one player's strategy uses \( (1-\delta)n \) random bits there exists
a deterministic strategy resulting in a payoff of \( \delta n \) against them.
This results in an exact characterization of approximate \nashes{} and required
randomness to play them in matching pennies.
Both of these results were later shown~\citep{DBLP:journals/mst/HubacekNU16} to
be consequences of~\citet{DBLP:journals/geb/NeymanO00}'s results, which
characterize the maxmin payoff of a player with bounded entropy in a repeated
two-player zero-sum game (against an opponent with access to unbounded entropy).
\citet{DBLP:journals/geb/GossnerV02} in turn generalize this result by
assuming that the player with bounded entropy only has access to realizations of
random variables \( X_t \sim \mathcal{L}(X) \) of entropy \( h =
\H(\mathcal{L}(X)) \) at each round~\( t \), whose \emph{arbitrary
distribution} \( \mathcal{L}(X) \) is also publicly known.
Follow-up work by~\citet{DBLP:journals/geb/ValizadehG19} further extends this
work to a setting where the source of entropy \( X \) can also be leaked to the
adversary, and study the non-asymptotic behavior of the maxmin
value~\citep{DBLP:journals/corr/abs-1902-03676}.
\citet{DBLP:conf/esa/KalyanaramanU07} undertake this problem with a learning
flavor: in two-player zero-sum games with payoffs in \( \{ 0, 1 \} \), they
provide an algorithm using \( O(\log \log n + \log(1/\varepsilon)) \) random
bits (with high probability) yielding a \( O \left( 1/\sqrt{n} + \varepsilon
\right) \) additive regret term against an adaptive adversary.
All of this work differs from our results in that it characterizes
off-equilibrium outcomes in games where players are bounded by their randomness,
whereas we characterize the amount of randomness players require to play
equilibria.

\paragraph{Computational \nashes{} with bounded randomness}
\citet{DBLP:conf/sigecom/BudinichF11} and~\citet{DBLP:journals/mst/HubacekNU16}
also study \emph{computational \nashes{}} in repeated games, the former for
matching pennies and the latter in two-player zero-sum games.
A computational \nashe{} is an approximate \nashe{} that can be implemented using
polynomial-size circuits.
\citet{DBLP:journals/mst/HubacekNU16} show that if one-way functions do not
exist, then there are no computational \nashes{} using sublinear entropy in
two-player zero-sum games with no weakly dominant pure strategies (recall that
they show there are no \nashes{} using constant entropy in the computationally
unbounded case).

\section{Preliminaries}\label{sec:prelim}
Let \( G \) be a normal-form game, with a set \( A = \llbracket 1; m
\rrbracket \) of \( m \geq 2
\) players, and denote \( S_i \) the set of strategies of player~\( i \) and \( S
= S_1 \times \cdots \times S_m \) the entire strategy space.
We assume that \( \card{S_i} \in \N^* \setminus \{ 1 \} \) for all \( i \).
\( \sigma \) will denote a (potentially mixed) strategy profile \( (\sigma_1,
\ldots, \sigma_m) \), where \( \sigma_i \) is the distribution with which player
\( i \) samples over \( S_i \).
Occasionally, when referring to a strategy profile \( \sigma_n \) that already
has a subscript, we will denote \( {\left( \sigma_n \right)}_i \) the strategy
profile of player~\( i \) in \( \sigma_n \).
When strategies are known to be pure (i.e.\ the actions are deterministic) we
will use the letter \( s \in S \).
We use the notation \( \sigma_{-i} \) to designate the strategy profile of all
players except player~\( i \) in \( \sigma \), i.e.\ \( (\sigma_1, \ldots,
\sigma_{i-1}, \sigma_{i+1}, \ldots, \sigma_m) \).
The utility of player~\( i \) is \( u_i: S \to [0; 1] \) (we assume all payoffs
are normalized without loss of generality).
This function is then extended to mixed strategies in the natural way,
\[
	u_i(\sigma) = \sum_{s \in S} \left( \prod_{j=1}^m \sigma_j(s) \right)
	u_i(s).
\]
We write \( u_i(\sigma_j, \sigma_{-j}) = u_i(\sigma) \), and \( u(s) = (u_1(s),
\ldots, u_m(s)) \).

\begin{defi}\label{defi:minmax}
	The \textbf{minmax} of player~\( i \) is the best payoff player~\( i \)
	can ensure under any play by the other players,
	\begin{equation}\label{eq:minmax}
		\tilde{v}_i = \min_{\sigma_{-i}} \; \max_{\sigma_i} \;
		u_i(\sigma) = \min_{\sigma_{-i}} \; \max_{s_i \in S_i} \;
		u_i(s_i, \sigma_{-i}).
	\end{equation}
	The \textbf{pure minmax} is the best payoff player~\( i \) can ensure
	assuming other players only play pure strategies,
	\begin{equation}\label{eq:pureminmax}
		v_i = \min_{s_{-i} \in \prod_{j \neq i} S_j} \; \max_{\sigma_i}
		\; u_i(\sigma_i, s_{-i})
		= \min_{s_{-i} \in \prod_{j \neq i} S_j} \; \max_{s_i \in S_i} \;
		u_i(s).
	\end{equation}
\end{defi}
Note the pure minmax of a player is always larger than their minmax.

\begin{defi}
	The set of \textbf{feasible payoff profiles} (or feasible payoffs) is the
	convex hull of \( \{ (u_1(s), \ldots, u_m(s)), \; s \in S \} \subset
	{[0;1]}^m \).
	We say that a particular feasible payoff profile \( p \) is
	\textbf{supported by} strategies \( (s_1, \ldots, s_r) \in S^r \) and
	coefficients \( (\gamma_1, \ldots, \gamma_r) \in {\left( \R^*_+
	\right)}^r \) if \( \sum_{k=1}^r \gamma_k u(s_k) = p \) and \(
	\sum_{k=1}^r \gamma_k = 1 \).
	We will also say that \( p \) is \textbf{supportable by} \( \{s_1,
	\ldots, s_r \} \subseteq S \) in this case.
\end{defi}
A payoff being feasible does not necessarily imply there exists a mixed strategy
\( \sigma \) achieving it, since mixed strategies only span product
distributions.
Note the difference with~\citet[Definition~4]{DBLP:journals/mst/HubacekNU16}'s
definition of feasible payoff profiles, where they only consider convex
combinations with coefficients in \( \Q \).
\begin{defi}\label{defi:qfeas}
	A feasible payoff profile is called \textbf{\( \Q \)-feasible} if it is
	supported by strategies and coefficients such that \( \gamma_k \in \Q \)
	for all \( k \) (when such an assumption is not made, we sometimes refer
	to the payoff profile as \( \R \)-feasible).

	A payoff profile \( p \) is \textbf{individually rational for player~\(
	i \)} if its payoff is larger than its minmax, \( p_i \geq \tilde{v}_i
	\), and \textbf{individually rational} if \( p \geq \tilde{v} \).
	Strictly individually rational means \( p > \tilde{v} \).

	\( p \) is \textbf{purely individually rational for player~\( i \)} if
	its payoff is larger than its pure minmax, \( p_i \geq v_i \) (and the
	previous variants extend).
\end{defi}

The extensive-form game where \( G \) is repeated \( n \in \N^* \) times is
denoted \( G^n \) and is called the \textbf{\( n \)-repeated game} of \( G \),
and \( G \) is called the \textbf{stage game} of \( G^n \).
A \textbf{history of play} during the first \( k \in \llbracket 1; n-1 \rrbracket
\) rounds is a \( k \)-tuple \( h \in S^k \).
When \( h \in S^{n-1} \) is a history of play for all first \( n-1 \) rounds,
subscripts denote its truncation to its first elements: \( h_k \) is the
history of play for the first \( k \) rounds of \( h \).
A strategy profile for \( G^n \) is a mapping from histories of play to mixed
strategy profiles of the stage game \( G \) for all players.
They are also denoted by the symbol \( \sigma \), and \( \sigma(h_k) \) refers to
the strategy profile of \( G \) played at round \( k+1 \) if the history of play
is \( h_k \) in the first \( k \) rounds.
The symbol \( \emptyset \) denotes the empty history, and \( \sigma(\emptyset) \)
is therefore the strategy profile played at the first round in \( \sigma \).
The concatenation of two histories \( h, h' \) is denoted \( h \cdot h' \).

\begin{defi}
	For a given strategy profile \( \sigma \) of a repeated game, we call the
	\textbf{on-path tree} \( T(\sigma) \) the tree of all histories that are
	sampled with non-zero probability by \( \sigma \): its root is \(
	\emptyset \) and the children of some history \( h \in S^k \) are the \(
	h \cdot s \in S^{k+1} \) such that \( s \) is played with non-zero
	probability in \( \sigma(h) \).
	We denote \( H(\sigma) \) the set of leaves of \( T(\sigma) \), and \(
	\P_\sigma(h) \) the probability of history \( h \in S^k \) when all
	players play according to \( \sigma \).
\end{defi}

Recall that for a discrete distribution \( p \) over a set \( S \), its
\textsc{Shannon} base-\( 2 \) entropy is defined as,
\[
	\H(p) = -\sum_{s \in S} p(s) \log_2(p(s)).
\]
\begin{defi}[{\citep[Definition~10]{DBLP:journals/mst/HubacekNU16}}]\label{defi:entropy}
	For a given strategy \( \sigma \) of the repeated game \( G^n \), we
	define its \textbf{entropy} (or total entropy) for player~\( i \) as,
	\[
		\H_i(\sigma) = \H_i(\sigma_i) = \max_{h \in S^{n-1}} \left[
			\H(\sigma_i(\emptyset)) + \sum_{k=1}^{n-1}
			\H(\sigma_i(h_k)) \right].
	\]
	Its \textbf{effective entropy} only considers histories that happen with
	non-zero probability during play,
	\[
		\Heff_i(\sigma) = \max_{h \in H(\sigma)} \left[
			\H(\sigma_i(\emptyset)) + \sum_{k=1}^{n-1}
			\H(\sigma_i(h_k)) \right].
	\]
\end{defi}

\begin{defi}
	The \textbf{amount of randomness} or \textbf{amount of entropy} of an
	equilibrium \( \sigma \) of a repeated game is \( \sum_{i=1}^m
	\H_i(\sigma) \).
\end{defi}

\subsection{Computational model}\label{sec:computational}
There are three immediate issues when the framework above is applied to
computationally-bounded players (for example \textsc{Turing}-equivalent players):
\renewcommand{\theenumi}{(\roman{enumi})}%
\begin{enumerate}
	\item All \nashes{} of the stage game \( G \) can have irrational
	coefficients in all generality (even when payoff matrices have
	coefficients in \( \Q \)) and cannot immediately be efficiently
	represented;~\footnote{However, since the set of NEs is a semialgebraic
	variety, the \textsc{Tarski}-\textsc{Seidenberg} principle ensures the
	existence of an efficiently representable NE satisfying any linear
	conditions on its payoffs provided there exists such a
	\nashe{}~\citep[Theorem~1]{DBLP:conf/latin/LiptonM04}.}
	\item\label{item:sampling} As a second consequence, it is not clear that
	sampling from a player's distribution for such an equilibrium can be done
	exactly (non-exact play could be exploited by other players) and;
	\item\label{item:overhead} Rather than sampling from one distribution
	with entropy \( \Heff_i(\sigma) \), a player might be sampling many times
	from conditionally independent very low-entropy distributions (at each
	round), which naïvely requires a much larger number of expected bits
	(because of overhead and uncertainty on what distributions it will have
	to sample from in the future).
\end{enumerate}
We therefore have to specify a computational model that addresses all of these
issues.

To represent \nashes{} or more generally any mixed strategy profile, we assume
that players have access to an oracle that can specify the probabilities of their
strategy up to any finite precision at each round.
A particular case of this model could be that agents have all approximately
computed the strategy profile themselves, and are able to compute approximations
of the distributions up to any precision efficiently.~\footnote{As per the
previous footnote, if all played \nashes{} (and strategy profiles) are algebraic
this would simply mean computing approximations to exactly-represented numbers.}

For point~\ref{item:sampling}, to sample from a \nashe{}
represented in this way, the players can use inversion sampling and request the
oracle for more bits as required.
Algorithm~\ref{alg:inversion} in Appendix~\ref{app:computational} contains an
inversion sampling algorithm for this oracle model, and
Proposition~\ref{prop:inversion} shows it requires a finite number of bits in
expectation (with a geometrically-decaying tail), a finite number of requests to
the oracle in expectation, and produces an exact
sample.

Regarding point~\ref{item:overhead}, throughout the paper we will ensure that
whenever there exists an equilibrium of a repeated game using \( O(1) \) entropy,
it will also use mixing in \( O(1) \) rounds along each path.
This ensures that it can be played with a constant expected amount of random bits
using the inversion sampling algorithm from point~\ref{item:sampling} at each
round.
Note that without this condition, using inversion sampling could require \(
\Omega(n) \) expected random bits to play some \( O(1) \)-entropy equilibria of
the repeated game due to overhead.

\begin{defi}
	Analogously to the definition of entropy of a repeated \nashe{} \(
	\H(\sigma) \), we define its \textbf{worst-case expected random
	bits}~\footnote{Note that this concept is distinct from the expected
	random bits conditioned on a worst-case path being taken for several
	reasons: (i) it could be ill-defined, as the worst-case path for player
	\( i \) could involve deviations by player~\( i \) and (ii) the
	definitions of total and effective entropy do not condition on the path
	being taken, and would be very different if they did.} for player~\( i \)
	as,
	\[
		\textsf{Bits}_i(\sigma) = \max_{h \in S^{n-1}} \min_N \; \E
		(N(\sigma_i(\emptyset), \sigma_i(h_1), \ldots, \sigma_i(h_n))),
	\]
	where \( N \) spans over the number of expected bits used by algorithms
	terminating with probability \( 1 \) and producing an exact sample of the
	distribution.
\end{defi}

Theorem~2.2 of~\citet{knuthyao76} and its corollary yield that,
\[
	\H_i(\sigma) \leq \textsf{Bits}_i(\sigma) < \H_i(\sigma) +2.
\]
This completes our computational model as it means that \nashes{} of the repeated
game requiring \( \Omega(n) \) entropy on their worst-case path will also induce
an algorithm using \( \Omega(n) \) bits in expectation at least along their
worst-case path.

\section{Two-player games}\label{sec:twoplayer}
In this section, we show that the necessary condition from
Proposition~\ref{prop:simplenecinf}, along with Lemma~\ref{lmm:necfeasibleinf}
are together sufficient in the two-player case, providing a complete
characterization of the existence of constant-entropy \nashes{} of the repeated
game.
The proof of the following theorem is deferred to Theorem~\ref{thm:twoplayerapp}
in Appendix~\ref{app:twoplayer}.
\begin{thm}\label{thm:twoplayer}
	A two-player game \( G \) has \( O(1) \)-randomness \nashes{} of its
	repeated game
	if and only if \( G \) has a pure \nashe{} or if it has a purely
	individually rational feasible payoff profile and has a \nashe{} that is
	strictly individually rational for a player.
\end{thm}

\paragraph{Sublinear randomness}
Notice that the proof of the necessary condition only relies on
Proposition~\ref{prop:simplenecinf} and Lemma~\ref{lmm:necfeasibleinf}, which
both only require sublinear-entropy \nashes{}.
We deduce that the same condition holds for sublinear-entropy \nashes{}, yielding
the following 0--1 law for entropy in two-player repeated games.
\begin{cor}
	For any two-player game, there are either \( O(1) \)-randomness
	equilibria of its repeated game or all equilibria require \( \Omega(n) \)
	randomness.
\end{cor}

\paragraph{Asymptotically achievable payoffs}
Another consequence of the proof is that by applying the necessary condition and
then the sufficient condition to a series of sublinear-entropy \nashes{}, we find
constant-entropy \nashes{} with the same asymptotic average payoffs.
This provides a folk theorem-like result for constant-entropy \nashes{}, which
can be compared to folk theorems such as~\citet{benoit1987nash}.
\begin{cor}\label{cor:payofftwo}
	When the condition of Theorem~\ref{thm:twoplayer} is satisfied, the set
	of payoff profiles achievable asymptotically by sublinear and
	constant-entropy equilibria are both exactly the set of purely
	individually rational payoff profiles.
	When the condition is not satisfied, there are no sublinear-entropy
	\nashes{} and the set of achievable payoffs with sublinear entropy is
	empty.
\end{cor}

\paragraph{Both players requiring \( \Omega(n) \) entropy each}
Theorem~\ref{thm:subsettwo} in Appendix~\ref{app:twoplayer} provides a
characterization for the existence of \nashes{} in which one player uses \( O(1)
\) entropy.
This yields the following characterization of two-player games in which all
\nashes{} of the repeated game require \( \Omega(n) \) entropy \emph{from both
players}.
Recall that~\citet{DBLP:journals/mst/HubacekNU16} proved that a sufficient
condition is that all \nashes{} of \( G \) have payoffs exactly \( \tilde{v} \)
(the minmax of all players) and all \nashes{} of \( G \) are mixed for all
players (for effective entropy).
We complete this into a necessary and sufficient condition for two-player games
using total entropy.
\begin{thm}\label{thm:omegatwo}
	Suppose \( G \) is a two-player game.
	All \nashes{} of the repeated game \( G^n \) require \( \Omega(n) \)
	randomness from both players if and only if all \nashes{} of \( G \) are
	mixed for all players and either all \nashes{} of \( G \) have payoffs
	exactly \( \tilde{v} \) or there is no individually rational feasible
	payoff profile \( p \) that is also purely individually rational for one
	of the two players.
\end{thm}

\section{Observable distributions}\label{sec:observable}
In this section, we prove a similar result for games with more than two players
in a setting where players are able to observe the distributions that other
players used to select their action at the end of each round.
This setting serves as an intermediate between the two-player case and the
general case, and the condition we obtain is a natural extension of
the one from Theorem~\ref{thm:twoplayer}.
The assumption of observable distributions is standard in the repeated game
literature~\citep{478688ac-522f-32d9-af21-b6cdd5ca9890,fudenberg1991dispensability}.
\begin{defi}\label{defi:observable}
	For a given game \( G \) and \( n \in \N^* \), \( \tilde{G}^n \) is the
	\textbf{\( n \)-repeated game with observable distributions} of \( G \).
	It is the extensive-form game where strategy profiles are mappings from
	histories of actions and  distributions for all players to strategy
	profiles of \( G \),
	\[ \sigma: \bigcup_{k=0}^{n-1} {\left(S \times \prod_{j \in A}
	\Delta(S_j) \right)}^k \to \prod_{j \in A} \Delta(S_j), \]
	where \( \Delta(X) \) is the simplex over \( X \).
\end{defi}
The main advantage of observable distributions is that it records any deviation
in the history, allowing for deviations in mixed rounds to be punished, which
cannot be done in the general case.
The proof of the following theorem is deferred to Theorem~\ref{thm:observableapp}
in Appendix~\ref{app:observable}.
\begin{thm}\label{thm:observable}
	An \( m \)-player game \( G \) has \( O(1) \)-randomness \nashes{} of its
	repeated game with observable distributions \( \tilde{G}^n \) (or
	equivalently sublinear-randomness \nashes{})
	if and only if it has a purely individually rational feasible payoff
	profile \( p \) supportable by \( S' \subseteq S \) and there exists a
	mapping \( f: A \to \{ -\infty \} \cup \N \) such that,
	\begin{itemize}
		\item If \( f(i) = -\infty \) then player~\( i \) best responds
		in all strategy profiles in \( S' \);
		\item If \( f(i) \in \N \) then there exists a strategy profile
		\( \sigma \) of \( G \) such that \( u_i(\sigma) >
		\tilde{v}_i \) and in which the players \( \{ j, \; f(j) \leq
		f(i) \} \) are all best responding.
	\end{itemize}
\end{thm}

\section{General case}\label{sec:general}
We finally extend our results to general repeated games (without observable
distributions).
The proof of the following theorem is deferred to
Theorem~\ref{thm:mostgeneralapp} in Appendix~\ref{app:mostgeneral}.
\begin{thm}\label{thm:mostgeneral}
	An \( m \)-player game \( G \) has \( O(1) \)-randomness \nashes{} of its
	repeated game \( G^n \)
	if and only if \( G \) has a feasible purely individually rational
	payoff profile \( p \) supportable by \( S' \subseteq S \), there exists
	some constant \( n_0 \in \N \), a \nashe{} \( \sigma_{n_0} \) of the
	repeated game \( G^{n_0} \), and a partition \( A = A_0 \cup A_1 \) of
	the players such that,
	\begin{itemize}
		\item Every player in \( A_0 \) is best responding in every
		strategy profile in \( S' \),
		\item Every player in \( A_1 \) has an average payoff in \(
		\sigma_{n_0} \) that is strictly better than their minmax.
	\end{itemize}
	Moreover, if \( G \) does not satisfy this condition then all equilibria
	of its repeated game require \( \Omega(n) \) entropy.
\end{thm}
As in Corollary~\ref{cor:payofftwo}, note the set of achievable payoffs by
constant or sublinear-entropy equilibria are the same, and are exactly the
payoff profiles \( p \) that satisfy the condition in the theorem above for some
support \( S' \), integer \( n_0 \), equilibrium \( \sigma_{n_0} \) and partition
\( A_0 \cup A_1 \).

The structure of the conditions from Theorems~\ref{thm:twoplayer}
and~\ref{thm:observable} is partially lost, as there is no ordering but simply a
partition: \( A_0 \) can be seen as \( f^{-1}(\{ -\infty \}) \) and \( A_1 \) as
\( f^{-1}(\N) \).
We now give a sufficient condition and a necessary condition that are closer to
our earlier characterizations.
The proof of the following proposition is deferred to
Proposition~\ref{prop:suffapp} in Appendix~\ref{app:mostgeneral}.
\begin{prop}[Sufficient condition]\label{prop:suff}
	If \( G \) has a feasible purely individually rational payoff profile \(
	p \) supportable by \( S' \subseteq S \) and there exists a mapping \( f:
	A \to \{ -\infty \} \cup \N \) such that,
	\begin{itemize}
		\item If \( f(i) = -\infty \) then player~\( i \) is best
		responding in every strategy profile in \( S' \)
		\item If \( f(i) \in \N \) then there exists a strategy profile
		\( \sigma_i \) such that \( u_i(s) > \tilde{v}_i \) and in which
		the players \( \{ j, \; f(j) \leq f(i) \} \) are best responding
		and the players \( \{ j, \; f(j) > f(i) \} \) have the same
		payoff for all the strategies they are mixing over,
	\end{itemize}
	then \( G \) has \( O(1) \)-randomness \nashes{} of its repeated game.
\end{prop}

\begin{cor}[Necessary condition]
	If \( G \) has \( o(n) \)-entropy \nashes{} of its repeated game, then
	the condition of Theorem~\ref{thm:observable} is verified by \( G \).
\end{cor}
\begin{proof}
	A \( o(n) \)-randomness equilibrium of the repeated game \( G^n \) is
	immediately a \( o(n) \)-randomness equilibrium of the repeated game with
	observable distributions \( \tilde{G}^n \).
	By Theorem~\ref{thm:observable}, it satisfies all of its conditions.
\end{proof}

\paragraph{Subset of players using bounded entropy}
Finally, we show that our techniques can be adapted to provide conditions on
games where only a subset of players use sublinear randomness.
\begin{defi}\label{defi:tindiv}
	For \( T \subseteq A \), a payoff profile \( p \) is called \textbf{\( T
	\)-individually rational} if,
	\[
		\forall i \in A, \qquad
		p_i \geq \min_{\substack{s_j \in S_j \\ j \in T \setminus \{ i \}}}
	 		 \;
			 \min_{\substack{\sigma_j \\ j \in A \setminus (T \cup \{ i \})}}
	 		 \;
			 \max_{s_i \in S_i}
			 u_i(s_i, s_{T \setminus \{ i \}},  \sigma_{A \setminus (
			 T \cup \{ i \})}) = v_{T,i}.
	\]
	The term on the right \( v_{T,i} \) is called the \textbf{\( T \)-pure
	minmax} of player~\( i \).
\end{defi}
In short, it must ensure the player's minmax assuming players from \( T \) do not
use mixing and those from \( A \setminus T \) do.
The proof of the following theorem is deferred to Theorem~\ref{thm:subsetapp} in
Appendix~\ref{app:subset}.
\begin{thm}\label{thm:subset}
	Suppose \( G \) is an \( m \)-player game, and \( T \subseteq A \).
	If \( G \) has a feasible \( T \)-individually rational payoff profile \(
	p \) supportable by \( S' \subseteq S \), there exists some constant \(
	n_0 \in \N \), a \nashe{} \( \sigma_{n_0} \) of the repeated game \(
	G^{n_0} \), and a partition \( A = A_0 \cup A_1 \) of the players such
	that,
	\begin{itemize}
		\item Every player in \( A_0 \) best responds in every strategy
		profile in \( S' \),
		\item Every player in \( A_1 \) has an average payoff in \(
		\sigma_{n_0} \) that is strictly better than their minmax,
	\end{itemize}
	then \( G \) has \nashes{} of its repeated game such that all players in
	\( T \) use \( O(1) \) randomness.

	Conversely, if \( G \) has \nashes{} of its repeated game
	such that all players in \( T \) use \( O(1) \) randomness, then it has a
	feasible \( T \)-individually rational payoff profile \( p \).
\end{thm}

\section*{Acknowledgements}
The author would like to thank Edan~Orzech and Kai~Jia for their feedback on the
computational model.
Laurel~Britt, Martin~Rinard, and Sandeep~Silwal are also thanked for their
feedback on early versions of the introduction.

\printbibliography{}

\newpage
\appendix
\section{Technical lemmas}
\begin{lmm}\label{lmm:tech1}
	For all \( x \in ]0; 1] \),
	\[ \min(x, 1-x) \leq -x \log_2(x). \]
\end{lmm}
\begin{proof}
	The function \( f: x \mapsto -x \log_2(x) \) (extended by continuity to
	\( [0; 1] \)) is concave, therefore it is above any line in its
	hypograph.
	Note that the points \( (0,0), (1/2,1/2) \) and \( (1, 0) \) all lie on
	the graph of \( f \).
	Therefore, we can conclude that \( f \) is above the lines connecting
	these points, which happen to exactly form the graph of \( x \mapsto
	\min(x, 1-x) \).
\end{proof}

\begin{lmm}\label{lmm:tech2}
	For all \( x \in [0; 1-1/e] \) and \( y \in \R_+ \),
	\[ {(1-x)}^y \geq 1 - 2xy. \]
\end{lmm}
\begin{proof}
	\[ {(1-x)}^y = \exp( \ln(1-x) y) \geq \exp(-2xy) \geq 1 - 2xy. \]
	The fact that \( \ln(1-x) \geq -2x \) is verified by concavity of \( x
	\mapsto \ln(1-x) \) and the fact that \( -2(1-1/e) < \ln(1/e) = -1 \)
	because \( e > 2 \).
	The fact that \( \exp(-t) \geq 1-t \) is verified by the fact that the
	exponential is convex and its tangent at \( 1 \) is \( t \mapsto 1 + t
	\).
\end{proof}

\begin{lmm}\label{lmm:xa}
	For any vector \( (x_1, \ldots, x_r) \in {\left(\R^*_+ \right)}^r \) such
	that \( \sum_{i=1}^r x_i = 1 \), there exists a sequence \( {\left( a_k
	\right)}_{k \in \N} \) such that \( a_k \in \N^r \) and \( \sum_{j=1}^r
	a_{k,j} = k \) and \( a_{k+1} \geq a_k \) and \( a_{k,j} \geq \lfloor k
	x_j \rfloor \) for all \( k \in \N \) and \( j \in \llbracket 1; r
	\rrbracket \).
\end{lmm}
\begin{proof}
	We define the series inductively as \( a_{0,j} = 0 \) for all \( j \),
	and \( a_{k+1} = a_k + \ind{j_0} \) where \( \ind{j_0} \) is the vector
	containing zeroes except a \( 1 \) at index \( j_0 \), and \( j_0 \) is
	defined as,
	\begin{equation}\label{eq:j0}
		j_0 = \arg\min_{j \in \llbracket 1; r \rrbracket}
			\left\lfloor \frac{a_{k,j} + 1}{x_j} \right\rfloor.
	\end{equation}
	Clearly, this construction ensures that \( \sum_{j=1}^r a_{k,j} = k \)
	and \( a_{k+1} \geq a_k \) for all \( k \).

	By contradiction, let \( k \in \N^* \) be the first index for which there
	is some \( j_1 \) such that \( a_{k,j_1} < \lfloor k x_{j_1} \rfloor \).
	In each round \( i \in \llbracket 1; k \rrbracket \), the vector \( a \)
	has been increased by \( 1 \) exactly.
	Moreover, the vector \( \lfloor ix \rfloor = \left( \lfloor ix_1 \rfloor,
	\ldots, \lfloor ix_r \rfloor \right) \) increases by at most \( i \) in
	the first~\( i \) rounds, since it is monotonous and its sum is at most
	\( i \) at round \( i \).

	For each round \( i \), let \( r(i) \) be the pair defined as,
	\[
		r(i) = \left( \min_{j \in \llbracket 1; r \rrbracket}
			\left\lfloor \frac{a_{i,j} + 1}{x_j} \right\rfloor, \;
			\arg \min_{j \in \llbracket 1; r \rrbracket}
			\left\lfloor \frac{a_{i,j} + 1}{x_j} \right\rfloor
			\right),
	\]
	with the same tie-breaking as in equation~\eqref{eq:j0} above.
	In words, it is the next round and index at which one of the constraints
	will be violated if the vector \( a_i \) is left unchanged.
	Its second component is also the index \( j_0 \) that is increased at
	step \( i+1 \).

	For a constraint \( a_{i,j} \geq \lfloor i x_j \rfloor \) to become first
	violated at round \( i \), it must be that \( \lfloor i x_j \rfloor =
	\lfloor (i-1) x_j \rfloor + 1 = a_{i,j} + 1 \).
	Increasing \( a_{i,j} \) by one ensures this constraint is no longer
	violated at round \( i \).
	The labels \( r(0), \ldots, r(k-1) \) therefore all point to \( k \)
	distinct increases by one of the vector \( \lfloor i x \rfloor \).
	Moreover, \( (k, j_1) \not\in \left\{ r(0), \ldots, r(k-1) \right\} \),
	since this constraint is violated at round \( k \).
	The labels \( r(i) \) are increasing in their first component and \(
	\left\lfloor \frac{a_{k-1,j_1} +1}{x_{j_1}} \right\rfloor = \left\lfloor
	\frac{a_{k,j_1} +1}{x_{j_1}} \right\rfloor = k \), meaning \( r(k-1) \)'s
	first component is at most \( k \).
	This is a contradiction, since it means there are \( k+1 \) increases in
	the vector \( \lfloor i x \rfloor \) within the first \( k \) rounds,
	labeled \( \{ r(0), \ldots, r(k-1), (k, j_1) \} \).
\end{proof}

\section{Computational model proofs}\label{app:computational}
We present an inversion sampling algorithm that can exactly sample from any given
distribution over a finite support when provided oracle access to any
finite-precision representation of the probabilities.
We assume that oracles return the unique representation of each \( p_i \) that
does not end with an infinitely repeating sequence \( 111\bar{1} \), and that all
of the \( p_i \) are strictly positive.

Note there are more efficient implementations possible of this algorithm's ideas,
but to simplify the proof we use a rather inefficient implementation.
\begin{algorithm}
\caption{Inversion sampling with access to binary approximation
	oracles}\label{alg:inversion}
\begin{algorithmic}
	\Require{} Oracle access to the probabilities \( (p_1, \ldots, p_n) \) up
	to any arbitrary precision
	\Ensure{} A sample \( X \) following the law, \( \P(X = i) = p_i \)
	\State{} \( k \gets 10 \)
	\State{} \( r \gets 0.\textsf{RandomBits}(10) \)
	\Comment{\parbox[t]{.6\linewidth}{A floating-point number with \( 10 \)
	random bits after the decimal point.}}
	\State{} \( b_0 \gets 0 \) and \( b_n \gets 0.111\bar{1} \)
	\While{True}
	\State{} \( q_i \gets \textsf{Oracle}(p_i, k) \) for \( i \in \llbracket
	1; n \rrbracket \) \Comment{The first \( k \) bits of \( p_i \).}
	\State{} \( b_i \gets \sum_{j=1}^i q_j \) for \( i \in \llbracket 1; n-1
	\rrbracket \)
	\If{\( r \not\in \cup_{i=1}^{n-1} [b_i;\, b_i+i2^{-k}[ \)}
		\State{} \Return{} \( i \) such that \( b_{i-1} \leq r \leq b_i
		\)
	\EndIf{}
	\State{} \( k \gets k + 1 \)
	\State{} \( r \gets r + 2^{-k} \times \textsf{RandomBits}(1) \)
	\Comment{Concatenate a new random bit.}
	\EndWhile{}
\end{algorithmic}
\end{algorithm}

\begin{prop}\label{prop:inversion}
	Algorithm~\ref{alg:inversion} generates a sample following the input
	distribution with probability \( 1 \), and uses a finite number of random
	bits and oracle calls in expectation.
\end{prop}
\begin{proof}
	Let \( R \sim \mathcal{U}([0;1]) \) be the random variable that would be
	obtained by continuously concatenating random bits at the end of \( r \).
	We will show that the algorithm returns \( i \) if and only if \( R \in
	] \sum_{j=1}^{i-1} p_j; \, \sum_{j=1}^i p_j [ \) conditionally to the
	event that \( R \) does not end with an infinite repeating series of \( 1
	\) and is not equal to any of the \( {\left( \sum_{j=1}^i p_j \right)}_{i
	\in \llbracket 0; n \rrbracket} \) (which happens with probability one),
	this will show the algorithm is exact.
	Fix some iteration \( k \in \llbracket 10; +\infty \llbracket \) and let
	\( r \) be the value of the variable \( r \) in
	Algorithm~\ref{alg:inversion} at this iteration.

	Note that by definition of the oracle, \( p_i \in [q_i; q_i + 2^{-k} [ \)
	at each loop.
	Note also that \( R \in [r, r+2^{-k}[ \) conditionally on the fact that
	the remaining bits are not all ones.
	This implies \( b_i \leq \sum_{j=1}^i p_i < b_i + i 2^{-k} \).
	Therefore, if \( r \geq b_i + i 2^{-k} \) then \( R \geq r > \sum_{j=1}^i
	p_i \).
	On the other hand, if \( r < b_i \), then \( r \leq b_i - 2^{-k} \) since
	both of these numbers have a binary representation in \( k \) bits.
	This means that \( R < r + 2^{-k} \leq \sum_{j=1}^i p_i \).
	Therefore, if the algorithm returns some index \( i \in \llbracket 2; \,
	n-1 \rrbracket \), since \( b_{i-1} < r < b_i \), it means that \( R \in
	] \sum_{j=1}^{i-1} p_j; \, \sum_{j=1}^i p_j [ \).
	When \( i = 1 \) or \( i = n \), the same analysis on the upper or lower
	boundary respectively and the conditioning \( R \not\in \{ 0; 1 \} \)
	yields the conclusion.

	Conversely, if \( R \in ] \sum_{j=1}^{i-1} p_j; \, \sum_{j=1}^i p_j [ \),
	there exists some large enough \( k \in \N^* \) such that,
	\[ \sum_{j=1}^{i-1} p_j + i 2^{-k} < R < \sum_{j=1}^i p_j - n 2^{-k}. \]
	At iteration \( k \), it holds that \( r > R - 2^{-k} \geq b_j + j 2^{-k}
	\) for all \( j < i \) and for all \( j \geq i \),
	\[ r \leq R <  \sum_{l=1}^j p_l - n 2^{-k} \leq b_j + j2^{-k} - n2^{-k}
	\leq b_j. \]
	This means the algorithm terminates at or before iteration \( k \).
	Since this is true for all \( i \in \llbracket 1; \, n \rrbracket \) and
	for almost all \( R \in [0;1] \), the algorithm terminates with
	probability one.
	Since the probability of the algorithm returning \( i \) is \(
	\sum_{j=1}^i p_j - \sum_{j=1}^{i-1} p_j = p_i \), it is also exact.

	We finally show the bound on the expected number of iterations, which
	implies the expected number of random bits sampled is also finite.
	If there are at least \( (k+1) \in \N^* \) iterations, it means that at
	iteration \( k \) the algorithm hasn't returned.
	There are \( 2^k \) possible values for \( r \) at iteration~\( k \), and
	the number of values that lead to non-termination are at most,
	\[
	\sum_{i=1}^{n-1} \card{\{ j2^{-k}, \, j \in \llbracket 0; 2^k-1
	\rrbracket \} \cap [b_i; b_i + i 2^{-k} [} =
	\sum_{i=1}^{n-1} i \leq n^2.
	\]
	This means the probability that the number of iterations is at least \( k
	\) is at most \( n^2 2^{-k} \).
	As the expectation of the number of iterations is the sum of the
	probabilities of the number of iterations being at least \( k \) for \( k
	\in \N^* \), we find that the expected number of iterations is bounded by
	\( 2 n^2 < +\infty \).
	Note that this also shows the number of calls to \textsf{Oracle} is
	finite in expectation, and that it (along with the number of sampled
	bits) decays at least geometrically.
\end{proof}

\section{Playing with sublinear randomness proofs}\label{app:sublinear}
\subsection{Tools for sufficient conditions}
\begin{prop}\label{prop:simplesuff}
	If \( G \) has a \( \Q \)-feasible payoff profile \( p \) that is purely
	individually rational for all players and every player has a \nashe{}
	that is strictly individually rational for them, then there exists
	a \nashe{} of the repeated game requiring \( O(1) \) entropy.
\end{prop}
\begin{proof}
	We build an equilibrium in two phases, one of variable length (which can
	be extended to any arbitrary length) and a second of constant length.

	Let \( \frac{1}{d} \sum_{k=1}^r a_k u(s_k) \) be a set of strategy
	profiles \( {\left( s_k \right)}_k \) and a convex combination supporting
	the payoff profile \( p \), where \( a_k \in \N^* \) for all \( k \).
	The first phase is as follows: play \( s_1 \) \( a_1 \) times, followed
	by \( s_2 \) \( a_2 \) times, etc\ldots{}
	Then repeat this cycle as many times as necessary, reaching any
	arbitrarily large number of rounds.
	If player~\( i \) deviates during any of these rounds, i.e.\ if the
	history contains a round where they played an action other than
	prescribed by the schedule above, play a pure minmax strategy profile
	for player~\( i \), i.e.\ a pure strategy profile \( s \) in the \(
	\arg\min \) and \( \arg\max \) from equation~\eqref{eq:pureminmax} (and
	maintain this play no matter what other deviations occur, meaning the
	rest of the history is disregarded for this entire phase).
	When the history contains deviations from several players, play any pure
	strategy profile \( s \in S \): these histories do not intervene in the
	definition of a \nashe{} and a pure strategy profile will avoid any
	contributions to the total entropy.

	The second phase is a constant number of rounds regardless of the number
	of repetitions of the payoff profile \( p \) in the first phase.
	Let \( \sigma_i \) be a strictly individually rational \nashe{} for each
	player~\( i \).
	Define \( \delta_i = u_i(\sigma_i) - \tilde{v}_i > 0 \) by definition of
	\( \sigma_i \).
	For each player in an arbitrary order, play \( \sigma_i \) for \(
	\left\lceil \frac{d}{\delta_i} \right\rceil \) rounds.
	This play ignores the second phase's history altogether: players keep
	playing the same sequence of \nashes{} no matter the history of play on
	the previous \nashes{}.
	The last thing that must be specified is what is played during this
	second phase of the game if there have been deviations during the first
	phase: if a single player~\( i \) deviated in the first phase, play the
	strategy profile in the \( \arg\min \) and \( \arg\max \) of
	equation~\eqref{eq:minmax} achieving its minmax \( \tilde{v}_i \) for the
	entire second phase.
	If several players deviated in the first phase, play any arbitrary pure
	strategy profile in \( S \) (this does not affect entropy or \nashe{}).

	The constructed strategy profile of the repeated game satisfies the
	constant entropy condition: the first phase consists of only pure
	strategies for all players (along any path) and the second phase is of
	length at most \( m + \sum_{i \in A} \frac{d}{\delta_i} =O(1) \) (and
	each round uses a finite amount of entropy in each path).
	It is left to show that it is indeed a \nashe{}.
	We claim that any player deviating during the first phase can gain at
	most \( d \) in expected payoff compared to their pure minmax over the
	entire first phase of play.
	Indeed, each round of the current repetition of \( p \) (including the
	round they deviated at) has payoff at most \( 1 \).
	We reason in terms of benefit from deviating, which is the difference
	between the expected payoff under any deviation and on-path play: in this
	case, it is at most \( 1-v_i \leq 1 \) per round for this repetition of
	\( p \), which means at most \( d \) in this repetition.
	However, following repetitions of \( p \) have an average payoff \( p_i
	\geq v_i \), therefore the rest of the first phase has benefit at most \(
	cd(v_i - p_i) \leq 0 \) where \( c \in \N \) is the remaining number of
	repetitions of \( p \) in the first phase, proving the claim.

	A player~\( i \) having deviated during the first phase will have their
	payoff during the rounds where their individually rational \nashe{} \(
	\sigma_i \) should have been played decreased.
	Their benefit is \( -\delta_i \) during at least \( \frac{d}{\delta_i} \)
	rounds, therefore at most \( -d \) over these rounds.
	During the rounds playing \( \sigma_j \) for any other player, their
	benefit is at most \( 0 \) since on-path play is a sequence of \nashes{},
	which by definition have a higher payoff than their minmax \( \tilde{v}_i
	\).
	This means that any deviation during the first phase benefits the
	deviating player at most \( d-d = 0 \), and therefore is not strictly
	profitable.
	During the second phase, any deviation is done during a \nashe{} and does
	not affect what will be played in future rounds, therefore no deviation
	is strictly profitable there either.
\end{proof}

In the case of \( \R \)-feasible payoff profiles, we need to use a completely
different construction, as a constant finite cycle length can no longer be
guaranteed.
We begin with this more general lemma which is used in the construction of the
equilibria.
\begin{lmm}\label{lmm:Rbound}
	If \( p \) is a \( \R \)-feasible payoff profile of \( G \), then for all
	\( n \in \N^* \) there exists a series of strategy profiles \( \left(
	s_1, \ldots, s_n \right) \) such that for all \( k \in \llbracket 1; n
	\rrbracket \), the average payoff of each player~\( i \) over the last \(
	k \) rounds is at least \( p_i - \card{S}/k \).
\end{lmm}
\begin{proof}
	Let \( \sum_{j=1}^r x_j u(y_j) = p \) be a support for \( p \), where \(
	y_j \in S \) for all \( j \).
	Build the series of integer vectors \( {\left( a_k \right)}_{k \in \N} \)
	according to the construction of Lemma~\ref{lmm:xa}, for the coefficients
	\( x = \gamma \).
	Define the series of strategies \( s_k = y_j \) where \( j \) is such
	that \( \ind{j} = a_{n-k+1} - a_{n-k} \).

	The strategies played in the last \( k \) rounds are \( s_{n-k+1},
	\ldots, s_n \), which contain strategies \( y_1, \ldots, y_r \) in
	proportions,
	\[
		a_k - a_{k-1} + a_{k-1} - a_{k-2} + \cdots + a_1 - a_0 = a_k.
	\]
	Therefore, using the notations from Lemma~\ref{lmm:xa}, the average
	payoff for the last \( k \) rounds is the matrix product,
	\[
		\frac{a_k}{k} \cdot
		\begin{pmatrix}
			u(y_1) \\
			\vdots \\
			u(y_r) \\
		\end{pmatrix} \geq
		\frac{\lfloor kx \rfloor}{k} \cdot
		\begin{pmatrix}
			u(y_1) \\
			\vdots \\
			u(y_r) \\
		\end{pmatrix} \geq
		\frac{kx - \mathbbm{1}}{k} \cdot
		\begin{pmatrix}
			u(y_1) \\
			\vdots \\
			u(y_r) \\
		\end{pmatrix} \geq
		p - \frac{r}{k} \mathbbm{1}. \qedhere
	\]
\end{proof}

\begin{prop}\label{prop:Rsuff}
	If \( G \) has an \( \R \)-feasible payoff profile \( p \) that is purely
	individually rational for all players and every player has a \nashe{}
	that is strictly individually rational for them, then there exists
	a \nashe{} of the repeated game requiring \( O(1) \) entropy.
\end{prop}
\begin{proof}
	The equilibrium used follows the same structure as the one from
	Proposition~\ref{prop:simplesuff}, except the first phase is replaced by
	the series from Lemma~\ref{lmm:Rbound}.
	The second phase consists of \( \left\lceil \frac{\card{S}+1}{\delta_i}
	\right\rceil \) repetitions of each player's strictly individually
	rational NE \( \sigma_i \), similarly to
	Proposition~\ref{prop:simplesuff}.
	The total entropy is still constant independently from the length of the
	first phase.

	It is left to show that it is indeed a \nashe{}.
	We claim that any player deviating during the first phase can gain at
	most \( \card{S} \) in expected payoff compared to their pure minmax over
	the entire first phase of play.
	Indeed, if player~\( l \) deviates \( k \) rounds before the end of the
	first phase, their maximum benefit over the rest of the phase is \( k v_l
	- k \left( p_l - \card{S}/k \right) \leq \card{S} \) by
	Lemma~\ref{lmm:Rbound}.
	Adding to this the benefit from deviating at the round where they
	deviate, their benefit is indeed upper bounded by \( \card{S}+1 \).
	Following the same proof as in Proposition~\ref{prop:simplesuff}, it
	follows that the second phase contains sufficient incentives to
	compensate this benefit, and therefore this is indeed a \nashe{}.
\end{proof}

We now use the proof of Proposition~\ref{prop:Rsuff} to prove the following
proposition, informally stated in the introduction.
\begin{prop}\label{prop:a0suff}
	If \( G \) has an \( \R \)-feasible purely individually rational payoff
	profile \( p \) supported by a convex combination of strategy profiles \(
	\sum_{k=1}^r \gamma_k s_k \) such that for each player~\( i \), either
	they have a strictly individually rational \nashe{} \( \sigma_i \) or
	they are best responding in all of the \( s_k \), then the repeated games
	\( G^n \) have \nashes{} requiring \( O(1) \) entropy.
\end{prop}
\begin{proof}
	Use the same equilibrium as in the proof of Proposition~\ref{prop:Rsuff},
	except the second phase only plays the \( \sigma_i \) for the players
	that have a strictly individually rational \nashe{}.
	A player who does not have a strictly individually rational \nashe{} is
	best responding at every round of both phases.
	Therefore they achieve their pure minmax at every round of the first
	phase and their minmax at every round of the second phase: any deviation
	reduces their payoff at every future round, including the one they
	deviated at.
\end{proof}

\subsection{Tools for necessary conditions}
We now prove a first necessary condition, which is similar to the contraposition
of Theorem~1 of~\citet{DBLP:journals/mst/HubacekNU16}, with the two differences
that it only requires no pure \nashes{} rather than all \nashes{} being mixed for
all players, and it requires \( o(n) \) randomness for all players rather than
for at least one player.
Our proof technique is different from theirs, and is reused in various necessary
conditions throughout the paper.
\begin{prop}\label{prop:simplenec}
	Assume \( G \) does not have a pure \nashe{}, and has \nashes{} of its
	repeated game using \( o(n) \) entropy.
	Then \( G \) has a \nashe{} that is strictly individually rational for at
	least one player.
\end{prop}
\begin{proof}
	First, note the set of \nashes{} of the game \( G \) is compact, and the
	mapping \( \H \) from strategy profile to used entropy (summed over all
	players) is continuous.
	This means that \( 0 \) is not an accumulation point for the entropies of
	\nashes{}, therefore there is some \( \delta > 0 \) such that all
	\nashes{} consume at least \( \delta \) entropy.
	Since every path in the on-path tree \( T(\sigma) \) uses sublinear
	entropy and each \nashe{} uses at least \( \delta > 0 \) entropy, for
	large enough \( n \) there is some node \( h \) in the on-path tree which
	is not a \nashe{} (in fact, there is such a node along every history in
	\( H(\sigma) \)).
	This is true because the entropy on a path grows sublinearly whereas for
	all nodes to be \nashes{} it would need to be at least \( \delta n \).
	Furthermore, this round cannot be the last round since the last round is
	always a \nashe{} in the on-path tree.

	Let \( i \) be a player that is not best responding in \( \sigma(h) \),
	and let \( k < n-1 \) be the length of \( h \) (i.e.\ it appears at depth
	\( k \) in the tree).
	We show there is at least one node in the subtree rooted at \( h \) in \(
	T(\sigma) \) that is a \nashe{} where player~\( i \) has payoff strictly
	greater than their minmax.
	First notice that all nodes in this subtree (except the root) are
	\nashes{}, since we have assumed the root of this subtree is at the last
	round where a player is not best responding.
	The expected payoff for player~\( 1 \) when playing \( \sigma \) after \(
	h \) is their payoff at the root, \( u_i(\sigma(h)) \), plus the
	expected payoff over all future rounds and possible action samplings of
	their payoffs.
	This means that their on-path payoff from \( h \) for the rest of the
	game can be written,
	\[
		u_i(\sigma(h)) + \sum_{j=k+1}^{n} \sum_{h'_j \in S^{j-k}}
		\P_\sigma(h \cdot h'_j \mid h) u_i(\sigma(h \cdot h'_j)),
	\]
	where \( h'_j \) are possible histories for the rest of the game after \(
	h \).
	On the other hand, best responding at the root and at all subsequent
	rounds will achieve expected payoff at least \( u_i(b, \sigma_{-i}(h)) +
	(n-k) \tilde{v}_i \) where \( b \in S_i \) is a best response to \(
	\sigma_{-i}(h) \) for player~\( i \).
	Since \( h \) is sampled with non-zero probability by definition (it is
	in the on-path tree), deviating at \( h \) is not strictly profitable,
	\[
		u_i(b, \sigma_{-i}(h)) + (n-k) \tilde{v}_i
		\leq
		u_i(\sigma(h)) + \sum_{j=k+1}^{n} \sum_{h'_j \in S^{j-k}}
		\P_\sigma(h \cdot h'_j \mid h) u_i(\sigma(h \cdot h'_j)).
	\]
	This in turn implies,
	\[
		0 < u_i(b, \sigma_{-i}(h)) - u_i(\sigma(h))
		\leq
		\sum_{j=k+1}^{n} \sum_{h'_j \in S^{j-k}}
		\P_\sigma(h \cdot h'_j \mid h) \left[ u_i(\sigma(h \cdot h'_j)) -
		\tilde{v}_i \right].
	\]
	Therefore, there exists some round \( j \) and history \( h'_j \) for
	which \( u(\sigma(h \cdot h'_j)) > \tilde{v}_i \) (since all rounds are
	NEs, all terms are greater than or equal to \( 0 \)) and \( \sigma(h
	\cdot h'_j) \) is a \nashe{} of \( G \).
\end{proof}

For the second necessary condition, we begin by lower bounding the worst payoff
that can be inflicted on a player by the other players with bounded entropy.
To do so, we begin by defining punishments.
\begin{defi}\label{defi:punish}
	An \( n \)-round punishment~\( \sigma \) of player~\( i \) in \( G \) is
	a strategy profile of the repeated game \( G^n \) such that player~\( i
	\) cannot benefit from deviating to any other \( \sigma_i' \), i.e.
	\begin{equation}\label{eq:punish}
		\forall \sigma_i', \; u_i(\sigma) \geq u_i(\sigma_i',
		\sigma_{-i}).
	\end{equation}
\end{defi}
\begin{lmm}\label{lmm:avg}
	For any small enough \( \varepsilon > 0 \) and any \( n \)-round
	punishment of player~\( i \) using at most \( O(f(n)) \) entropy the
	average payoff of player~\( i \) is at least
	\[
		v_i - O \left( \frac{f(n)}{\varepsilon n} \right) -
		O(\varepsilon).
	\]
\end{lmm}
\begin{proof}
	Let \( \sigma \) be a worst punishment.
	We can assume without loss of generality that player~\( i \) best
	responds at each node in \( T(\sigma) \), since inductively no matter
	what \( i \) does the worst punishment must be played in the rest of the
	punishment.
	More formally, we can prove by induction that there is a worst punishment
	\( \sigma \) where player~\( i \) is purely best responding at every node
	of \( T(\sigma) \).
	It is clear for a one-round punishment: \( i \) is best responding and
	without loss of generality we can replace that best response by a pure
	best response.
	For any given worst \( n \)-round punishment \( \sigma \), replace every
	subtree of the root of \( T(\sigma) \) with a worst \( (n-1) \)-round
	punishment where \( i \) purely best responds at every on-path node, and
	replace the strategy of player~\( i \) at the root of \( T(\sigma) \) by
	an arbitrary best response \( b \) to the other player's strategies \(
	\sigma_{-i}(\emptyset) \).
	This is a punishment for player~\( i \), since no deviation at the root
	increases their payoff and every subtree of the root is a punishment by
	assumption.
	It is also a worse punishment than \( \sigma \): the payoff of player~\(
	i \) at the root is the expected payoff of playing \( b \) plus the
	expected payoff of the \( (n-1) \)-round worst punishment.
	This is smaller than the expected payoff of playing \( b \) plus the
	expected payoff of the \( (n-1) \)-round punishments prescribed by \(
	\sigma \) when \( (b, \sigma_{-i}(\emptyset)) \) are played at the root.
	This in turn is smaller by the punishment condition of
	equation~\eqref{eq:punish} than the expected payoff of playing \(
	\sigma(\emptyset) \) at the root.
	This therefore is an \( n \)-round worst punishment where player~\( i \)
	purely best responds at every node of the on-path tree.

	The expected payoff for the punishment can be re-written as a weighted
	average of paths from the on-path tree,
	\[
		\sum_{h \in H(\sigma)} \P_\sigma(h) \sum_{j=1}^n
		u_i(\sigma(h_j)).
	\]
	We will lower bound the payoff along each of these paths -- the inner
	sum -- effectively bounding the total expected payoff.

	Each path uses at most \( O(f(n)) \) entropy: fix some \( \varepsilon > 0
	\), consider rounds that use entropy at least \( \varepsilon \) and those
	that don't.
	Those that do will achieve payoff at least \( \tilde{v}_i \) (the minmax
	of player~\( i \), since player~\( i \) is best responding) and there are
	at most \( O \left( \frac{f(n)}{\varepsilon} \right) \) of these.

	On the other hand, consider a round where the entropy used is at most \(
	\varepsilon \), call \( \sigma \) its strategy profile.
	This means that for each player~\( j \) their distribution \( {\left(
	\sigma_j(s_j) \right)}_{s_j \in S_j} \) satisfies,
	\[
		\H(\sigma) = \sum_{s_j \in S_j} (-\sigma_j(s_j)
		\log_2(\sigma_j(s_j)))
		\leq \varepsilon.
	\]
	Using Lemma~\ref{lmm:tech1} this means that for every player~\( j \in A
	\) and for every action \( a \in S_j \), it holds that \(
	\min(\sigma_j(a), 1-\sigma_j(a)) \leq \varepsilon \) and therefore \(
	\sigma_j(a) \leq \varepsilon \) or \( \sigma_j(a) \geq 1 - \varepsilon
	\).

	Consider \( \bar{\sigma} \) the rounded strategy profile, where each of
	these probabilities \( \sigma_j(a) \) is replaced by \( 0 \) or by \( 1
	\) depending on which it is closest to.
	The expected payoff for player~\( i \) can be bounded by,
	\[
		\sum_{s \in S} \left( \prod_{j \in A} \sigma_j(s_j) \right)
		u_i(s)
		\geq
		{(1-\varepsilon)}^m \sum_{s \in S} \left( \prod_{j \in A}
		\bar{\sigma}_j(s_j) \right) u_i(s)
		\geq {(1-\varepsilon)}^m v_i.
	\]
	Indeed, the term that is rounded to \( 1 \) will have probability at
	least \( {(1 - \varepsilon)}^m \)
	and its payoff (which is the payoff of a pure strategy profile in which
	player~\( i \) is best responding) is at least \( v_i \) for player~\( i
	\).
	We can then use Lemma~\ref{lmm:tech2} for small enough \( \varepsilon \)
	to lower bound it as,
	\[
		\sum_{s \in S} \left( \prod_j \sigma_j(s_j) \right) u_i(s)
		\geq v_i \left( 1 - 2 m \varepsilon \right).
	\]

	\textit{In fine}, the average payoff along each path in this punishment
	for player~\( i \) is at least,
	\begin{align*}
		& O \left( \frac{f(n)}{n \varepsilon} \right) \tilde{v}_i +
		\left( 1 - O \left( \frac{f(n)}{\varepsilon n} \right) \right)
		v_i \left( 1 - 2 m \varepsilon \right)
		\\
		& \qquad \qquad = v_i \left( 1 - O \left( \frac{f(n)}{\varepsilon
		n} \right) \right) \left( 1 - O \left( \varepsilon \right)
		\right) - O \left( \frac{f(n)}{\varepsilon n} \right) \\
		& \qquad \qquad = v_i - O \left( \frac{f(n)}{\varepsilon n}
		\right) - O(\varepsilon).
		\qedhere
	\end{align*}
\end{proof}

\begin{lmm}\label{lmm:necfeasible}
	If \( G \) has a \nashe{} of its repeated game for all large enough \(
	n \) using \( o(n) \) randomness, then \( G \) has a \( \R \)-feasible
	payoff profile that is \textbf{purely} individually rational for all
	players.
\end{lmm}
\begin{proof}
	Consider player~\( i \) at round \( 1 \): their expected payoff for the
	entire game is greater than their expected payoff when deviating
	(including the deviation gain or loss at the first round, at most \( 1 \)
	in absolute value, therefore \( O(1/n) \) in terms of average payoff).
	When player~\( i \) deviates at the root, the subtree that is played has
	expected payoff greater than the worst punishment: without loss of
	generality, player~\( i \) deviates to a strategy that maximizes its
	payoff over these \( n-1 \) rounds hence it is a punishment (see
	Definition~\ref{defi:punish}).

	Let \( f(n) = o(n) \) be the entropy used by the \nashes{} of the
	repeated game.
	By Lemma~\ref{lmm:avg} in equation~\eqref{eq:worstpunish} the average
	payoff for the punishment is at least \( v_i - O \left( \varepsilon
	\right) - O \left( \frac{f(n)}{n \varepsilon} \right) \).

	On the other hand, the average payoff on-path is some feasible payoff
	vector \( u_n \) for each \( n \), averaging over all rounds, all
	possible histories and the probabilities of playing each strategy at each
	round according to each history,
	\begin{equation}\label{eq:wholeavg}
		u_n = \sum_{s \in S} \left[
			\sum_{h \in H(\sigma_n)} \P_{\sigma_n}(h)
			\frac{1}{n} \sum_{t=1}^n
			\left( \prod_{j \in A}
			{\left(\sigma_n\right)}_j(h_t)(s_j) \right) \right]
			u(s).
	\end{equation}
	For each \( n \), let \( \gamma_n \) be the probability distribution over
	\( S \) in equation~\eqref{eq:wholeavg}, such that \( u_n = \sum_{s \in
	S} \gamma_{n,s} u(s) \).
	As \( n \) goes to infinity, the coefficients \( {\left( \gamma_{n,s}
	\right)}_{s \in S} \) have an accumulation point in the compact set of
	probability distributions over \( S \).
	We can therefore extract a subseries of the series of \nashes{} whose
	distributions over payoffs \( \gamma_n \) converge to an element of the
	simplex over \( S \), and therefore the payoff profiles \( u_n \)
	converge to a limit feasible payoff vector \( u = \sum_{s \in S} \gamma_s
	u(s) \).

	Choose some arbitrarily small \( \varepsilon > 0 \), then choose \( n_0
	\) large enough such that for all \( n \geq n_0 \), the average payoff
	profile \( u_n \) of the \nashe{} of \( G^n \) verifies \( {\left\| u -
	u_n \right\|}_\infty \leq \varepsilon \).
	By \nashe{} at the first round for each player~\( i \), the average
	payoffs satisfy,
	\[
		v_i \leq u_i + \varepsilon + O \left( \frac{f(n)}{n \varepsilon}
		\right) + O \left( \frac{1}{n} \right).
	\]
	Since \( f(n) = o(n) \) and this is true for all large enough \( n \) and
	all \( \varepsilon > 0 \), we find \( v \leq u \), which concludes the
	proof.
\end{proof}

\section{Two-player games proofs}\label{app:twoplayer}
\subsection{Constant-entropy characterization}
\begin{thm}\label{thm:twoplayerapp}
	A two-player game \( G \) has \( O(1) \)-randomness \nashes{} of its
	repeated game
	if and only if \( G \) has a pure \nashe{} or if it has a purely
	individually rational feasible payoff profile and has a \nashe{} that is
	strictly individually rational for a player.
\end{thm}
\begin{proof}
	Begin by the sufficient condition, and assume \( G \) has no pure
	\nashes{}.
	We build an equilibrium in two phases, much like
	Proposition~\ref{prop:Rsuff}, except we need to distinguish between two
	cases.
	Up to renaming, assume player~\( 2 \) is the one that has a strictly
	individually rational \nashe{} \( \sigma_2 \), and denote \( p \) the
	purely individually rational payoff profile.

	The first case is when the maximum payoff of player~\( 1 \) is exactly
	their minmax \( \tilde{v}_1 \).
	Let \( p = \sum_{k=1}^r \gamma_k u(s_k) \) be a set of strategies
	supporting \( p \), the purely individually rational payoff profile.
	Then we have,
	\[
		\tilde{v}_1 \leq v_1 \leq
		\sum_{k=1}^r \gamma_k u_1(s_k) \leq
		\sum_{k=1}^r \gamma_k \tilde{v}_1 = \tilde{v}_1.
	\]
	Since for all \( k \) we have \( u_1(s_k) \leq \tilde{v}_1 \), this means
	the \( u_1(s_k) \) are all equal to \( \tilde{v}_1 = v_1 \).
	On the other hand, \( p_2 = \sum_k \gamma_k u_2(s_k) \geq v_2 \).
	Choose some \( k \in \llbracket 1; r \rrbracket \) such that \( u_2(s_k)
	\geq v_2 \).
	The pure strategy profile \( s_k \) is purely individually rational for
	both players, and achieves the maximum payoff of player~\( 1 \).
	Similarly to Proposition~\ref{prop:simplesuff}, the equilibrium we build
	is a first phase where only \( s_k \) is played for arbitrarily many
	rounds, and a second phase where \( \sigma_2 \) is played a constant
	number of rounds to reward or punish any deviations by player~\( 2 \).
	Let \( \delta_2 = u_2(\sigma_2) - \tilde{v}_2 > 0 \), the equilibrium \(
	\sigma_2 \) is played for \( k_2 = \left\lceil \frac{1}{\delta_2}
	\right\rceil \) rounds.
	Deviations by player~\( 1 \) are completely ignored for the entire game,
	whereas for player~\( 2 \) deviations during the first phase are punished
	by their pure minmax \( v_2 \) for the rest of the first phase and their
	minmax \( \tilde{v}_2 \) during the second phase.
	Deviations in the second phase are ignored.

	If player~\( 2 \) deviates during the first phase they benefit by at most
	\( 1 \) at the round they deviate, and for the rest of the first phase
	they benefit at most \( 0 \) because \( s_k \) is purely individually
	rational for them.
	During the second phase they receive \( \tilde{v}_2 \) at each round
	instead of \( u_2(\sigma_2) \), for \( k_2 \) rounds: their overall
	benefit from deviating is at most \( 1-\delta_2 k_2 \leq 0 \).
	Player~\( 1 \) on the other hand is always receiving their maximum payoff
	during the first phase, and is playing \nashes{} during the second phase,
	thus clearly has no strictly profitable deviations.

	\bigskip{}
	In the second case, we know the maximum payoff of player~\( 1 \) is
	strictly greater than their minmax \( \tilde{v}_1 \): let \( s \in S \)
	be a pure strategy profile achieving player~\( 1 \)'s maximum payoff.
	The first phase is the same as in the proof of
	Proposition~\ref{prop:Rsuff}, by using Lemma~\ref{lmm:Rbound} to produce
	a first phase of arbitrary length with average payoff at least \( p -
	\card{S}/k \) for the last \( k \) rounds, for all \( k \).

	Let \( \delta_1 = u_1(s) - \tilde{v}_1 > 0 \) by assumption.
	The second phase begins with \( k_1 = \left\lceil
	\frac{\card{S}+1}{\delta_1} \right\rceil \) repetitions of \( s \),
	followed by \( k_2 = \left\lceil \frac{\card{S}+1+k_1}{\delta_2}
	\right\rceil \) repetitions of \( \sigma_2 \).
	As in Proposition~\ref{prop:Rsuff}, any deviation in the first phase is
	punished by that player's pure minmax for the rest of the phase and that
	player's minmax during the entire second phase.
	Moreover, any deviation by player~\( 2 \) during the play of \( s \) is
	punished by playing the minmax of player~\( 2 \) for the rest of the
	game.
	Deviations during the play of \( \sigma_2 \) are ignored, since it is
	already a \nashe{}, and deviations by player~\( 1 \) during \( s \) are
	also ignored since it is already its maximum payoff.

	This clearly is a constant-entropy strategy profile of the repeated game,
	since randomness is only used in at most \( k_1 + k_2 = O(1) \) rounds
	along any path.
	It is left to show it is a \nashe{}.
	If player~\( 1 \) deviates during the first phase of the game they
	benefit at most \( \card{S}+1 \) in payoff, by Lemma~\ref{lmm:Rbound}:
	their benefit is at most \( 1 \) the round where they deviate, and over
	the rest of the first phase it is at most \( k v_1 - k(p_1 - \card{S}/k)
	\leq \card{S} \) where \( k \) is the remaining number of rounds in the
	first phase.
	However, doing so reduces their payoff in the second phase by at least \(
	\delta_1 \frac{\card{S}+1}{\delta_1} = \card{S}+1 \), meaning the
	deviation is not strictly profitable.
	Since \( s \) is already the maximum payoff for player~\( 1 \), they
	clearly have no interest in deviating from it, and they also have no
	interest in deviating during the play of \nashe{} \( \sigma_2 \) (in both
	cases their deviations have no effect on future rounds).

	Similarly, player~\( 2 \) can benefit by at most \( \card{S}+1 \) over
	the first phase of play by deviating during the first phase.
	Note that it is possible that \( u_2(s) < \tilde{v}_2 \), since player~\(
	2 \) is not necessarily best responding, so deviating during the rounds
	playing \( s \) could benefit them up to \( k_1 \).
	Therefore, any deviation at any round by player~\( 2 \) benefits them up
	to \( \card{S} + 1 + k_1 \) in payoff, however all of these deviations
	would be punished during the play of \( \sigma_2 \), leading to a loss of
	at least \( -\delta_2 \frac{\card{S} + 1+k_1}{\delta_2} = -\card{S} -
	1-k_1 \): no deviation is strictly profitable.

	\bigskip{}
	Conversely, assume that there are \( O(1) \)-randomness \nashes{} for
	arbitrary repetitions of the game, and that \( G \) has no pure
	\nashes{}.
	By Proposition~\ref{prop:simplenec}, there is a \nashe{} of \( G \) that
	is strictly individually rational for at least one player.
	Finally, Lemma~\ref{lmm:necfeasible} provides a feasible purely
	individually rational payoff profile \( p \) of \( G \).
\end{proof}

\subsection{Only one player using \( O(1) \) entropy}
\begin{thm}\label{thm:subsettwo}
	In a two-player game \( G \), there are \nashes{} of the repeated game in
	which player~\( 1 \) uses \( O(1) \) randomness if and only if either
	there is a \nashe{} of \( G \) where player~\( 1 \) doesn't mix, or there
	is an individually rational payoff profile \( p \) that is purely
	individually rational for player~\( 2 \) and there is a \nashe{} of \( G
	\) that is strictly individually rational for a player.
\end{thm}
\begin{proof}
	The sufficient condition is clear when there is a \nashe{} in which
	player~\( 1 \) does not mix.
	Let player~\( j \in \{ 1, 2 \} \) be the one that has a strictly
	individually rational \nashe{} \( \sigma_j \), and let \( i = 3-j \) be
	the other player.
	We distinguish the same two cases as in Theorem~\ref{thm:twoplayerapp}.

	The first case is when the maximum payoff of player~\( i \) is exactly
	their minmax \( \tilde{v}_i \).
	Let \( p = \sum_{k=1}^r \gamma_k u(s_k) \) be a set of strategies
	supporting \( p \), the purely individually rational payoff profile.
	Let \( \hat{v}_1 = \tilde{v}_1 \) and \( \hat{v}_2 = v_2 \), we have
	assumed that \( p \geq \hat{v} \).
	Then we have,
	\[
		\tilde{v}_i \leq \hat{v}_i \leq
		\sum_{k=1}^r \gamma_k u_i(s_k) \leq
		\sum_{k=1}^r \gamma_k \tilde{v}_i = \tilde{v}_i.
	\]
	Since for all \( k \) we have \( u_i(s_k) \leq \tilde{v}_i \), this means
	the \( u_i(s_k) \) are all equal to \( \tilde{v}_i \).
	On the other hand, \( p_j = \sum_k \gamma_k u_j(s_k) \geq \hat{v}_j \).
	Choose some \( k \in \llbracket 1; r \rrbracket \) such that \( u_2(s_k)
	\geq \hat{v}_2 \).
	The pure strategy profile \( s_k \) satisfies \( u(s_k) \geq \hat{v} \)
	like \( p \), and achieves the maximum payoff of player~\( i \).
	Like in Theorem~\ref{thm:twoplayerapp}, the equilibrium we build
	is a first phase where only \( s_k \) is played for arbitrarily many
	rounds, and a second phase where \( \sigma_j \) is played a constant
	number of rounds.
	Let \( \delta = u_j(\sigma_j) - u_j(\tilde{v}_j) > 0 \), we play \(
	\sigma_j \) for \( k = \left\lceil \frac{1}{\delta} \right\rceil \)
	rounds.
	Deviations by player~\( i \) are completely ignored for the entire game,
	whereas for player~\( j \) deviations during the first phase are punished
	according to,
	\begin{itemize}
		\item If \( j=1 \), they are punished by their minmax \(
		\tilde{v}_1 \) for any deviation for the rest of the game;
		\item If \( j=2 \), during the first phase they are punished by
		their pure minmax \( v_2 \) for the rest of the first phase and
		their minmax \( \tilde{v}_2 \) during the second phase.
	\end{itemize}
	Notice that this means that player~\(j \)'s punishment is \( \hat{v}_j \)
	for the rest of the first phase when they deviate.
	Deviations in the second phase are ignored.

	If player~\( j \) deviates during the first phase they benefit at most \(
	1 \) at the round they deviate, and for the rest of the first phase they
	benefit at most \( 0 \) because \( u_j(s_k) \) is at least \( \hat{v}_j
	\), which is also their punishment at each round.
	During the second phase they receive \( \tilde{v}_j \) at each round
	instead of \( u_j(\sigma_j) \), for \( k \) rounds: their overall
	benefit from deviating is at most \( 1-\delta k \leq 0 \).
	Player~\( i \) on the other hand is always receiving their maximum payoff
	during the first phase, and is playing \nashes{} during the second phase,
	thus clearly has no strictly profitable deviations.

	Finally, the amount of entropy used by player~\( 1 \) is at most
	constant, since the second phase is of length \( O(1) \) and the first
	phase only punishes player~\( 2 \) with their pure minmax.

	\bigskip{}
	In the second case, we know the maximum payoff of player~\( i \) is
	strictly greater than their minmax \( \tilde{v}_i \): let \( s \in S \)
	be a pure strategy profile achieving player~\( i \)'s maximum payoff.
	The first phase is the same as in the proof of
	Proposition~\ref{prop:Rsuff}, by using Lemma~\ref{lmm:Rbound} to produce
	a first phase with average payoff at least \( p - \card{S}/k \) for the
	last \( k \) rounds, for all \( k \).
	Let \( \delta' = u_i(s) - \tilde{v}_i > 0 \).
	The second phase begins with \( k_i = \left\lceil
	\frac{\card{S}+1}{\delta'} \right\rceil \) repetitions of \( s \),
	followed by \( k_j = \left\lceil \frac{\card{S}+1+k_i}{\delta}
	\right\rceil \) repetitions of \( \sigma_j \).

	Player~\( 1 \)'s deviations in the first phase are punished by their
	minmax \( \tilde{v}_1 \) for the rest of the game, whereas player~\( 2
	\)'s deviations during any point of the game are punished by their pure
	minmax \( v_2 \) for the rest of the first phase (if it is not over) and
	their minmax \( \tilde{v}_2 \) during the second phase.
	This means for the first phase the punishment of players are exactly \(
	\hat{v} \).
	Note that player~\( 1 \) uses constant entropy: in the first phase they
	play pure strategies on-path and pure punishments if player~\( 2 \)
	deviates, and the second phase is of length \( O(1) \).

	Finally, it is indeed a \nashe{} because either player deviating in the
	first phase benefits by at most \( \card{S}+1 \), and player~\( j \)
	gains at most \( k_i \) for deviating during the rounds rewarding
	player~\( i \).
	This is true because the first phase punishments are exactly \( \hat{v}
	\) and \( p \geq \hat{v} \) by assumption.
	We then see that \( k_i \) and \( k_j \) are set sufficiently large to
	punish such deviations.

	\bigskip{}
	For the necessary condition, assume there is a series of \nashes{} using
	sublinear entropy and assume player~\( 1 \) mixes in all \nashes{} of the
	game.
	Lemma~\ref{lmm:avgsubset} provides a payoff profile \( p \) satisfying
	the conditions above (note that \( T \)-individually rational for
	player~\( i \) reduces to individually rational or purely individually
	rational depending on whether the player~\( i \) is the one in \( T \) or
	not).
	The mapping from a mixed strategy profile to the entropy used by
	player~\( 1 \) only is continuous and the simplex is compact, therefore
	\( 0 \) cannot be an accumulation point, hence all \nashes{} require
	entropy at least \( \delta > 0 \) from player~\( 1 \) for some \( \delta
	\).
	In the same way as in Proposition~\ref{prop:simplenec}, for large enough
	\( n \) there is an on-path round in which a player is not best
	responding.
	By taking the latest such node in the on-path tree, its subtree must
	contain a \nashe{} that is strictly individually rational for the player
	that was not best responding at that node by the same argument as in
	Proposition~\ref{prop:simplenec}.
\end{proof}

\section{Observable distributions proof}\label{app:observable}
\begin{thm}\label{thm:observableapp}
	An \( m \)-player game \( G \) has \( O(1) \)-randomness \nashes{} of its
	repeated game with observable distributions \( \tilde{G}^n \) (or
	equivalently sublinear-randomness \nashes{})
	if and only if it has a purely individually rational feasible payoff
	profile \( p \) supportable by \( S' \subseteq S \) and there exists a
	mapping \( f: A \to \{ -\infty \} \cup \N \) such that,
	\begin{itemize}
		\item If \( f(i) = -\infty \) then player~\( i \) best responds
		in all strategy profiles in \( S' \);
		\item If \( f(i) \in \N \) then there exists a strategy profile
		\( \sigma_i \) of \( G \) such that \( u_i(\sigma_i) >
		\tilde{v}_i \) and in which the players \( \{ j, \; f(j) \leq
		f(i) \} \) are all best responding.
	\end{itemize}
\end{thm}
\begin{proof}
	We build a two-phase equilibrium like in the previous theorems: the first
	phase follows the construction of Lemma~\ref{lmm:Rbound}, and then add a
	constant amount of rounds needed to further reward or punish players.

	Order the players in \( f^{-1}(\N) \) by their images by \( f \),
	arbitrarily breaking ties when they have the same image.
	Label them \( i_0, i_1, \ldots, i_r \) such that \( 0 \leq f(i_0) \leq
	f(i_1) \leq \cdots \leq f(i_r) \), and define \( \delta_i = u_i(\sigma_i)
	- \tilde{v}_i > 0 \) for each of these players.
	For the first player~\( i_0 \), add as many rounds of \( \sigma_{i_0} \)
	needed to punish any gain they might have had from deviating in the
	first phase.
	As in previous proofs, \( k_0 = \left\lceil
	\frac{\card{S}+1}{\delta_{i_0}} \right\rceil \) repetitions are
	sufficient.
	For any deviation by any player during this punishment, play the minmax
	of that player for the rest of the game, and ignore multiple deviations.
	Note that deviations during \( \sigma_{i_0} \) are detectable even when
	it is a mixed profile, since distributions are observable after each
	round.
	Moreover, any future deviation by player~\( i_0 \) does not need to be
	punished, since we will only be playing strategy profiles in which they
	are best responding.
	The rest of play therefore ignores any deviation by \( i_0 \) in the
	second phase.

	Then, do the same for the next player~\( i_1 \) with its profile \(
	\sigma_{i_1} \), taking into account the maximum possible gain of \(
	\card{S}+1+k_0 \) it could have received by deviating during the first
	phase and \( k_0 \) by deviating during the punishment of player~\( i_0
	\).
	Repeating this procedure we find a number of repetitions \( k_j \) of \(
	\sigma_{i_j} \) sufficient to punish player~\( i_j \) is inductively
	defined by,
	\begin{equation}\label{eq:punishk}
		k_j = \left\lceil \frac{\card{S} + 1 + \sum_{j'=0}^{j-1}
		k_{j'}}{\delta_{i_j}} \right\rceil.
	\end{equation}
	Similarly, deviations by player~\( i_1 \) during \( \sigma_{i_1} \) or
	any future round are ignored.

	Note that the players with maximal \( f \)-value must be punished by
	\nashes{} that are strictly better than their pure minmax, ensuring the
	game ends with a \nashe{} -- and recovers the condition of
	Proposition~\ref{prop:simplenecinf}.
	If all players map to \( -\infty \) then the game is already all
	pure \nashes{} and we recover the first alternative in the condition for
	Theorem~\ref{thm:twoplayer}.

	We have added at most \( \sum_{j=1}^r k_j = O(1) \) rounds using entropy
	on any path, ensuring the amount of randomness is at most constant
	regardless of the length of the first phase.
	We are left to show this is a \nashe{} of the repeated game.
	Consider some player~\( i \), if \( f(i) = -\infty \) then that player is
	best responding during all of the game.
	Moreover, the punishment has worse payoff than the expected on-path
	payoff at each round: during the first phase pure strategy profiles in
	which \( i \) is best responding are played, which have payoff at least
	\( v_i \) by definition; during the second phase mixed strategy profiles
	in which \( i \) is best responding are played, which have payoff at
	least \( \tilde{v}_i \) by definition.

	Now if \( f(i) \in \N \), let \( i = i_j \) for some \( j \).
	Before \( \sigma_i \) is reached, the first phase is played as well as
	punishments for players ordered before \( i \).
	Any deviation by \( i \) is detectable and will benefit them at most \(
	\card{S} + 1 + \sum_{j'=0}^{j-1} k_{j'} \): if they deviate \( l \)
	rounds before the end of the first phase they gain at most \( 1 + lv_i -
	l(p-\card{S}/l) \leq \card{S} + 1 \), and then at most \( 1 \) at each
	round punishing other players before \( \sigma_i \).
	Deviating in \( \sigma_i \) or rounds following \( \sigma_i \) is not
	beneficial, since player~\( i \) is best responding and their deviations
	are ignored.
	Therefore, since punishing player~\( i \) by replacing \( \sigma_i \) by
	\( \tilde{v}_i \) reduces their payoff by \( \delta_i \), the number of
	rounds in equation~\eqref{eq:punishk} is sufficient to compensate the
	benefit of any previous deviation.

	\bigskip{}
	Conversely, for the necessary condition, Lemma~\ref{lmm:necfeasible}
	provides a feasible purely individually rational payoff profile \( p \),
	and let \( \sum_{s \in S'} \gamma_s u(s) \) be the coefficients and
	strategies \( S' \) to which on-path play converges.
	Note that the proof of Lemma~\ref{lmm:necfeasible} also applies in the
	case of repeated games with observable distributions with no
	modifications.
	Let \( f(i) \) be, for player~\( i \), the latest round in which it is
	not best responding at any node in the on-path tree \( T(\sigma) \),
	taking it to be \( -\infty \) if the player is always best responding.
	More precisely, since orderings of the players are a finite set (we can
	reduce the mapping from \( \{ -\infty \} \cup \N \) to \( \{ -\infty \}
	\cup \llbracket 1; m \rrbracket \) while preserving all conditions), one
	appears an infinite number of times, let \( f \) be such a mapping.
	Extract a subseries of the \nashes{} for which the ordering is always the
	one of \( f \), it is left to show that \( f \) satisfies the conditions
	of the theorem.

	If \( f(i) = -\infty \), we must show player~\( i \) best responds in
	every strategy supported by \( p \) (i.e.\ those in \( S' \)).
	Note that there is no guarantee that any particular strategy in \( S' \)
	will be played, since a little entropy could be used to mix away from it.
	We know that for all \( s \in S \),
	\[
		\sum_{h \in H(\sigma_n)} \P_{\sigma_n}(h) \frac{1}{n}
		\sum_{t=1}^n \left( \prod_{j \in A}
		{\left(\sigma_n\right)}_j(h_t)(s_j) \right) \xrightarrow[n \to
		+\infty]{} \gamma_s,
	\]
	by definition of \( p \) and \( \gamma_s \) (see the proof of
	Proposition~\ref{lmm:necfeasible}).
	This means there exists some \( n_0 \in \N \) such that,
	\begin{equation}\label{eq:existsh}
		\forall n \geq n_0, \; \forall s \in S', \; \exists h \in
		H(\sigma_n), \quad
		\sum_{t=1}^n \prod_{j \in A}
		{\left(\sigma_n\right)}_j(h_t)(s_j)
		\geq n \frac{\gamma_s}{2}.
	\end{equation}
	Let \( s \in S' \) be some strategy profile in the support of \( p \), we
	wish to show that \( i \) is best responding in it.
	Let \( g(n) = o(n) \) be the amount of entropy being used by \( \sigma_n
	\).
	Set some small \( 0 < \varepsilon < \gamma_s/4 \) and choose some \( n
	\geq n_0 \) such that \( n \frac{\gamma_s}{2} > 2 n\varepsilon +
	\frac{g(n)}{\varepsilon} \).
	Fix some history \( h \) according to equation~\eqref{eq:existsh}.
	Similarly to the proof of Lemma~\ref{lmm:avg}, we consider rounds that
	use at least \( \varepsilon \) entropy, and note there are at most \(
	\frac{g(n)}{\varepsilon} \) of them.
	Subtracting from both sides in equation~\eqref{eq:existsh} the
	contribution from these rounds (which is at most \( 1 \)),
	\[
		\sum_{t=1}^n \ind{\left( \sum_{i \in A}
		\H({\left(\sigma_n\right)}_i) \leq \varepsilon \right)} \left(
		\prod_{j \in A} {\left(\sigma_n\right)}_j(h_t)(s_j) \right) \geq
		n \frac{\gamma_s}{2} - \frac{g(n)}{\varepsilon} > 2 n
		\varepsilon.
	\]
	Therefore, there exists some round \( t \in \llbracket 1;n \rrbracket \)
	and history \( h \in H(\sigma_n) \) such that \( \sigma(h_t) \) uses
	entropy at most \( \varepsilon \), and for which the quantity in the
	left-hand sum is at least \( 2 \varepsilon \).
	Let \( \sigma = \sigma_n(h_t) \), this means,
	\begin{equation}\label{eq:mixbound}
		\prod_{j \in A} \sigma_j(s) \geq 2 \varepsilon.
	\end{equation}
	As seen in the proof of Lemma~\ref{lmm:avg}, a strategy profile using
	entropy at most \( \varepsilon \) for player~\( i \) must satisfy either
	\( \sigma_i(a) \leq \varepsilon \) or \( \sigma_i(a) \geq 1 - \varepsilon
	\) for all actions \( a \in S_i \).
	If any \( \sigma_j(s) \leq \varepsilon \) then
	equation~\eqref{eq:mixbound} is violated, therefore \( \sigma_j(s) \geq 1
	- \varepsilon \) for all \( j \in A \).

	We have thus shown that there is an on-path round in which all players
	are sampling \( s \) with probability at least \( 1-\varepsilon \) and in
	which player~\( i \) is best responding: \( \forall s_i' \in S_i, \;
	u_i(s_i', \sigma_{-i}) \geq u_i(\sigma) \).
	Both sides are continuous functions of \( \sigma \), and our property
	above is true for all small enough \( \varepsilon > 0 \), therefore by
	taking the limit we find that \( i \) is best responding in \( s \): \(
	\forall s_i' \in S_i, \; u_i(s_i', s_{-i}) \geq u_i(s) \).

	Finally, if \( f(i) \in \N \), there is at least one round in the on-path
	tree where player~\( i \) is not best responding for each \( \sigma_n \).
	By the same reasoning as in Proposition~\ref{prop:simplenec}, we find
	that there is at least one node in the subtree rooted at \( f(i) \) that
	has payoff strictly better than its minmax.
	By definition of \( f \), all players \( j \) such that \( f(j) \leq f(i)
	\) are best responding at that node, therefore \( \sigma_i \) can be
	chosen as the strategy profile at that node.
\end{proof}

\section{General case proofs}\label{app:general}
\subsection{Conditions for \( O(1) \)-entropy equilibria}\label{app:mostgeneral}
We finally show our most general result, holding for \( m \)-player games without
observable distributions.
\begin{thm}\label{thm:mostgeneralapp}
	An \( m \)-player game \( G \) has \( O(1) \)-randomness \nashes{} of its
	repeated game \( G^n \)
	if and only if \( G \) has a feasible purely individually rational
	payoff profile \( p \) supportable by \( S' \subseteq S \), there exists
	some constant \( n_0 \in \N \), a \nashe{} \( \sigma_{n_0} \) of the
	repeated game \( G^{n_0} \), and a partition \( A = A_0 \cup A_1 \) of
	the players such that,
	\begin{itemize}
		\item Every player in \( A_0 \) best responds in every strategy
		profile in \( S' \),
		\item Every player in \( A_1 \) has an average payoff in \(
		\sigma_{n_0} \) that is strictly better than their minmax.
	\end{itemize}
	Moreover, if \( G \) does not satisfy this condition then all equilibria
	of its repeated game require \( \Omega(n) \) random bits.
\end{thm}
\begin{proof}
	Begin with the sufficient condition, the equilibrium we build follows the
	same construction as previous proofs: a first phase using
	Lemma~\ref{lmm:Rbound}, and a second phase to punish or reward players.
	The second phase is a constant number of independent repetitions of the
	equilibrium \( \sigma_{n_0} \), where each repetition is played according
	to the equilibrium \( \sigma_{n_0} \), ignoring history from past
	repetitions of the equilibrium.

	Let \( \delta = \min_{i \in A_1} (u_i(\sigma_{n_0}) - n_0 \tilde{v}_i) \)
	which is strictly positive by definition.
	The number of repetitions of \( \sigma_{n_0} \) is \( \left\lceil
	\frac{\card{S}+1}{\delta} \right\rceil \).
	Deviations during the first phase of play are punished by the pure minmax
	of the deviating player until we reach the final phase, in which their
	minmax is played.

	Players in \( A_0 \) have no interest in deviating in the first phase, as
	shown in Proposition~\ref{prop:a0suff}.
	Players in \( A_1 \) deviating in the first stage will benefit in a
	payoff increase of at most \( \card{S}+1 \), but the second phase will
	punish them by at least \( -\delta_i \frac{\card{S}+1}{\delta} \leq
	-\card{S}-1 \) by definition.

	For the second phase, we show as a more general claim that the
	independent concatenation of a \nashe{} \( \sigma_n \) of the repeated
	game \( G^n \) and a \nashe{} \( \sigma_m \) of \( G^m \) form a \nashe{}
	\( \sigma_{n+m} \) of the repeated game \( G^{n+m} \).
	Indeed, assume there is a deviation that is strictly profitable to a
	player~\( i \) in \( \sigma_{n+m} \).
	If the first deviation from on-path play occurs in the latter \( m \)
	rounds, then the restriction of that strategy profile of player~\( i \)
	to the last \( m \) rounds is a strictly profitable deviation in \(
	\sigma_m \) (since \( \sigma_m \) is played regardless of history in the
	first \( n \) rounds), which is a contradiction.
	Otherwise, compare the expected payoff of the deviation in the first \( n
	\) and the last \( m \) rounds: their sum is strictly positive, therefore
	one of the terms is strictly positive.
	If it is the latter then proceed as above, if it is the former then that
	same deviation in \( \sigma_n \) is strictly profitable, another
	contradiction.

	By repeating this argument \( \left\lceil \frac{\card{S}+1}{\delta}
	\right\rceil \) times, we obtain that no unilateral deviation in the
	second phase is strictly profitable for any of the players.
	Finally, this \nashe{} uses \( O(1) \) total entropy, since only the
	second phase uses mixing and is of length \( O(1) \).

	\bigskip{}
	For the necessary condition, assume sublinear randomness is used in the
	\nashes{}, in order to prove both statements of the theorem at once.
	First use Lemma~\ref{lmm:necfeasible} to find the \( \R \)-feasible
	individually rational payoff profile \( p \) to which the on-path play of
	the series converges (assume without loss of generality we extract for
	that), and let \( S' \) be the support it converges to.
	For each \( n \) let \( A_{0,n} \) be players that are best responding at
	every on-path node of \( \sigma_n \) and \( A_{1,n} = A \setminus A_{0,n}
	\).
	The set of partitions of \( A \) is finite, therefore there is a
	partition appearing an infinite number of times: extract a subseries so
	that it is always the same, and name it \( A_0, A_1 \).
	By the same proof as Theorem~\ref{thm:observableapp} (an equilibrium of
	the non-observable version is a particular equilibrium of the observable
	version), any player in \( A_0 \) is best responding in every strategy
	profile in \( S' \).

	Now let \( i \) be some player in \( A_1 \), denote \( h \) a lowest
	node in \( T(\sigma) \) at which player~\( i \) is not best responding
	(for some fixed \( n \), after the previous extractions).
	All of the nodes in the subtree rooted at \( h \) have payoff at least \(
	\tilde{v}_i \) for player~\( i \), since they are by definition best
	responding at them all.
	However, as seen in the proof of Proposition~\ref{prop:simplenec}, there
	exists a node in the subtree that has payoff strictly better than \(
	\tilde{v}_i \) (and this node cannot be the root).
	Call the history of that node \( h' \), \( k_i \)
	the length of \( h' \) (i.e.\ the round number minus one, or its depth in
	the tree), and call \( \sigma_i \) the equilibrium of \(
	G^{n-k_i} \) that is prescribed by \( \sigma \) at \( h' \).

	We then let \( n_0 = \sum_{i \in A_1} (n-k_i) \), and define the
	equilibrium \( \sigma_{n_0} \) as the independent concatenation of the \(
	\sigma_i \) (as defined above).
	As seen above in the sufficient condition, this is a \nashe{} of \(
	G^{n_0} \).
	Moreover, each player~\( i \)'s average payoff in \( \sigma_i \) is
	strictly greater than their minmax.
	Finally, for \( j, j' \in A_1 \) player~\( j \)'s average payoff in \(
	\sigma_{j'} \) is at least \( \tilde{v}_j \), otherwise they
	could profitably deviate at the first round of \( \sigma_{j'} \) (which
	is a \nashe{}) and ensure an average payoff of \( \tilde{v}_j \).
	Thus the average payoff of every player in \( A_1 \) in \( \sigma_{n_0}
	\) is strictly greater than their minmax.
	This proves the necessary condition is true, which in turn also proves
	that \( O(1) \)-randomness equilibria of the repeated game exist,
	concluding our proof.
\end{proof}

\begin{prop}[Sufficient condition]\label{prop:suffapp}
	If \( G \) has a feasible purely individually rational payoff profile \(
	p \) supportable by \( S' \subseteq S \) and there exists a mapping \( f:
	A \to \{ -\infty \} \cup \N \) such that,
	\begin{itemize}
		\item If \( f(i) = -\infty \) then player~\( i \) best responds
		in every strategy profile in \( S' \)
		\item If \( f(i) \in \N \) then there exists a strategy profile
		\( \sigma_i \) such that \( u_i(s) > \tilde{v}_i \) and in which
		the players \( \{ j, \; f(j) \leq f(i) \} \) are best responding
		and the players \( \{ j, \; f(j) > f(i) \} \) have the same
		payoff for all the strategies they are mixing over,
	\end{itemize}
	then \( G \) has \( O(1) \)-randomness \nashes{} of its repeated game.
\end{prop}
\begin{proof}
	Build the same strategy profile of the repeated game as in
	Theorem~\ref{thm:observableapp}, except punish deviations differently
	during play of the \( \sigma_i \) (as deviations during mixed play are
	not observable in the history).
	Ignore all second-phase history, except when a player~\( i \) during a
	round \( \sigma_j \) plays an action outside of their support in \(
	{\left( \sigma_j \right)}_i \), and only when \( f(j) < f(i) \).
	In that case, \( \sigma_i \) is yet to be played: replace all rounds of
	\( \sigma_i \) by the minmax of player~\( i \).
	Deviations during the first phase are punished in the same way as in
	Theorem~\ref{thm:observableapp} and are therefore not strictly
	profitable.
	We are left to show that despite deviations not perfectly being
	punished anymore during the second phase, no player has a strictly
	profitable deviation.

	Consider some player~\( i \in f^{-1}(\N) \), if they deviate during a
	round playing \( \sigma_j \) such that \( f(j) \geq f(i) \) then their
	payoff is not strictly increased (since they are already best responding
	and their deviation is ignored).
	If they deviate during a round playing \( \sigma_j \) such that \( f(j) <
	f(i) \), then all the strategies they are mixing over have the same
	expected payoff over the rest of the game.
	Let \( S_i' \subseteq S_i \) be the support of \( {(\sigma_j)}_i \),
	there are two types of deviations for player~\( i \): those that preserve
	the support and those that extend it.
	If it preserves the support then the payoff at this round does not
	change, and the expected payoff over future rounds does not change (since
	by definition future play is unaffected by the outcome of this round if
	it remains in the same support) therefore the deviation is not strictly
	profitable.
	If it doesn't preserve the support, there is some \( a \in S_i \setminus
	S_i' \) that has non-zero probability in the deviation.
	The expected payoff conditional on playing \( a \) is at most increased
	by \( \sum_{j, \; f(j) < f(i)} k_j \) (using notations of
	Theorem~\ref{thm:observableapp}), however it will also trigger punishment
	during \( \sigma_i \), and we have ensured that \( k_i \) is large enough
	to punish such a profit in equation~\eqref{eq:punishk}.
	The expected payoff of playing \( a \) is therefore strictly smaller than
	the expected payoff of each action in \( S_i' \): this deviation strictly
	decreases \( i \)'s expected payoff.
\end{proof}

\subsection{Extension to effective entropy}\label{app:effective}
We show here a version of Theorem~\ref{thm:mostgeneral} using effective entropy
rather than total entropy (see Definition~\ref{defi:entropy}), to match the
setting of~\citet{DBLP:journals/mst/HubacekNU16}.
We begin by adapting Lemma~\ref{lmm:necfeasible} to this setting.
\begin{lmm}\label{lmm:necfeasibleeffective}
	If \( G \) has \nashes{} of its repeated game using \( O(1) \) effective
	entropy, then \( G \) has a \( \R \)-feasible payoff profile that is
	individually rational for all players.
\end{lmm}
\begin{proof}
	In any punishment of player~\( i \), they must have payoff at least equal
	to their minmax at each round, as shown in the proof of
	Lemma~\ref{lmm:avg}.
	The average payoff on-path is some feasible payoff vector \( u_n \) for
	each \( n \), by averaging over all rounds, all possible histories and
	all probabilities at each round as in equation~\eqref{eq:wholeavg}.
	By using the same reasoning as in the proof of
	Lemma~\ref{lmm:necfeasible}, after extracting a convergent sub-series in
	the coefficients \( {\left( \gamma_{n,s} \right)}_{s \in S} \), and
	choosing a small enough \( \varepsilon \), by \nashe{} at the first round
	for each player~\( i \), \( \tilde{v}_i \leq u_i + \varepsilon. \)
\end{proof}

\begin{thm}\label{thm:effective}
	Suppose \( G \) is an \( m \)-player game.
	\( G \) has \nashes{} of its repeated game using \( O(1) \) effective
	entropy
	if and only if it has a feasible individually rational payoff profile \(
	p \) supportable by \( S' \subseteq S \), there exists some constant \(
	n_0 \in \N \), a \nashe{} \( \sigma_{n_0} \) of the repeated game \(
	G^{n_0} \), and a partition \( A = A_0 \cup A_1 \) of the players such
	that,
	\begin{itemize}
		\item Every player in \( A_0 \) best responds in every strategy
		profile in \( S' \),
		\item Every player in \( A_1 \) has an average payoff in \(
		\sigma_{n_0} \) that is strictly better than their minmax.
	\end{itemize}
\end{thm}
\begin{proof}
	The sufficient condition follows from the same proof as
	Theorem~\ref{thm:mostgeneralapp}, where punishments for deviations in the
	first phase are replaced by the minmax of the deviating player (rather
	than their pure minmax).
	The final phase is constantly many independent repetitions of \(
	\sigma_{n_0} \) to punish any deviation.
	This uses constant effective entropy, since entropy is used on-path only
	during the second phase, which is of length \( O(1) \).
	It is a \nashe{} by the same proof as in Theorem~\ref{thm:mostgeneralapp}
	except both the lower bound on \( p \) and the punishments are replaced
	by \( \tilde{v} \) instead of \( v \).

	The necessary condition follows the same path as in
	Theorem~\ref{thm:mostgeneralapp}, using
	Lemma~\ref{lmm:necfeasibleeffective} above instead of
	Lemma~\ref{lmm:necfeasible}.
	All arguments used in Theorem~\ref{thm:mostgeneralapp} to construct the
	partition \( A_0 \cup A_1 \) and the equilibrium \( \sigma_{n_0} \) only
	use the fact that entropy used on-path is sublinear.
\end{proof}

\subsection{Playing with sublinear entropy for a subset of
players}\label{app:subset}
In this section, we show how our techniques extend to bounding entropy used by
only a subset \( T \subseteq A \) of players, by studying games with equilibria
\( \sigma_n \) such that \( \H_T(\sigma_n) = \sum_{i \in T} \H_i(\sigma_n) = O(1)
\).
Recall that in the two-player case, a complete characterization was provided in
Theorem~\ref{thm:subsettwo} in Appendix~\ref{app:twoplayer}.
We begin by recalling the definition of individually rational payoffs when only
part of the players are mixing.
\begin{defi}
	A payoff profile \( p \) is called \textbf{\( T \)-individually rational}
	if,
	\[
		\forall i \in A, \qquad
		p_i \geq \min_{\substack{s_j \in S_j \\ j \in T \setminus \{ i \}}}
	 		 \;
			 \min_{\substack{\sigma_j \\ j \in A \setminus (T \cup \{ i \})}}
	 		 \;
			 \max_{s_i \in S_i}
			 u_i(s_i, s_{T \setminus \{ i \}},  \sigma_{A \setminus (
			 T \cup \{ i \})}) = v_{T,i}.
	\]
	The term on the right \( v_{T,i} \) is called the \textbf{\( T \)-pure
	minmax} of player~\( i \).
\end{defi}
In short, it must ensure the players' minmax assuming players from \( T \) do not
use mixing and those from \( A \setminus T \) do.
Next, we show an adaptation of Lemma~\ref{lmm:avg} to the setting where only a
subset of players are bounded in randomness.
\begin{lmm}\label{lmm:avgsubset}
	For any small enough \( \varepsilon > 0 \) and any \( n \)-round
	punishment of player~\( i \) where players in \( T \) use at most \(
	O(f(n)) \) entropy the average payoff of player~\( i \) is at least
	\[
		v_{T,i} - O \left( \frac{f(n)}{\varepsilon n} \right) -
		O(\varepsilon).
	\]
\end{lmm}
\begin{proof}
	Let \( \sigma \) be a worst punishment, as shown in
	Lemma~\ref{lmm:avg} we can assume without loss of generality that
	player~\( i \) best responds at each node in \( T(\sigma) \).

	The expected payoff for the punishment can be re-written as a weighted
	average of paths from the on-path tree,
	\[
		\sum_{h \in H(\sigma)} \P_\sigma(h) \sum_{j=1}^n
		u_i(\sigma(h_j)).
	\]
	We will lower bound the payoff along each of these paths -- the inner
	sum -- effectively bounding the total expected payoff.

	Along each path, the sum of entropy used by players in \( T \) is \(
	O(f(n)) \): fix some \( \varepsilon > 0 \), consider rounds where their
	entropy \( \H_T(\sigma) \) is at least \( \varepsilon \) and those where
	it isn't.
	Those where it is will achieve payoff at least \( \tilde{v}_i \) (the
	minmax of player~\( i \), since player~\( i \) is best responding) and
	there are at most \( O \left( \frac{f(n)}{\varepsilon} \right) \) of
	these.

	On the other hand, consider a round where the entropy used by players in
	\( T \) is at most \( \varepsilon \), call \( \sigma \) its strategy
	profile.
	This means that for each player \( j \in T \) their distribution \(
	{\left( \sigma_j(a) \right)}_{a \in S_j} \) satisfies,
	\[
		\H(\sigma) = \sum_{a \in S_j} (-\sigma_j(a) \log_2(\sigma_j(a)))
		\leq \varepsilon.
	\]
	Using Lemma~\ref{lmm:tech1} this means that for every player \( j \in T
	\) and for every action \( a \in S_j \), it holds that \(
	\min(\sigma_j(a), 1-\sigma_j(a)) \leq \varepsilon \) and therefore \(
	\sigma_j(a) \leq \varepsilon \) or \( \sigma_j(a) \geq 1 - \varepsilon
	\).
	Let \( \bar{s}_T \) be the pure strategy profile in \( \prod_{j \in T}
	S_j \) for which \( \sigma_j(\bar{s}_{T,j}) \geq 1-\varepsilon \) for
	each \( j \in T \).
	The expected payoff for player~\( i \) being punished can be bounded by,
	\[
		\sum_{s \in S} \left( \prod_{j \in A} \sigma_j(s_j) \right)
		u_i(s)
		\geq
		{(1-\varepsilon)}^{\card{T}} \sum_{\substack{s \in S \\ s_T =
		\bar{s}_T}}
		\left( \prod_{j \in A \setminus T} \sigma_j(s_j) \right) u_i(s)
		\geq {(1-\varepsilon)}^{\card{T}} v_{T,i}.
	\]
	The last inequality holds because its left-hand side sum is the payoff of
	a strategy profile where player~\( i \) is best responding and the
	players mixing are at most \( A \setminus T \), and by
	Definition~\ref{defi:tindiv} of the \( T \)-individually rational payoff
	profile \( v_{T,i} \).

	We can then use Lemma~\ref{lmm:tech2} for small enough \( \varepsilon \)
	to lower bound it as,
	\[
		\sum_{s \in S} \left( \prod_{j \in A} \sigma_j(s_j) \right)
		u_i(s) \geq v_{T,i} \left( 1 - 2 \card{T} \varepsilon \right).
	\]

	\textit{In fine}, the average payoff along each path in this punishment
	for player~\( i \) is at least,
	\begin{align*}
		& O \left( \frac{f(n)}{n \varepsilon} \right) \tilde{v}_i +
		\left( 1 - O \left( \frac{f(n)}{\varepsilon n} \right) \right)
		v_{T,i} \left( 1 - 2 \card{T} \varepsilon \right)
		\\
		& \qquad \qquad = v_{T,i} - O \left( \frac{f(n)}{\varepsilon n}
		\right) - O(\varepsilon).
		\qedhere
	\end{align*}
\end{proof}
\begin{thm}\label{thm:subsetapp}
	Suppose \( G \) is an \( m \)-player game, and \( T \subseteq A \).
	If \( G \) has a feasible \( T \)-individually rational payoff profile \(
	p \) supportable by \( S' \subseteq S \), there exists some constant \(
	n_0 \in \N \), a \nashe{} \( \sigma_{n_0} \) of the repeated game \(
	G^{n_0} \), and a partition \( A = A_0 \cup A_1 \) of the players such
	that,
	\begin{itemize}
		\item Every player in \( A_0 \) best responds in every strategy
		profile in \( S' \),
		\item Every player in \( A_1 \) has an average payoff in \(
		\sigma_{n_0} \) that is strictly better than their minmax,
	\end{itemize}
	then \( G \) has \nashes{} of its repeated game such that all players in
	\( T \) use \( O(1) \) randomness.

	Conversely, if \( G \) has \nashes{} of its repeated game
	such that all players in \( T \) use \( O(1) \) randomness, then it has a
	feasible \( T \)-individually rational payoff profile \( p \).
\end{thm}
\begin{proof}
	The sufficient condition follows the same proof as
	Theorem~\ref{thm:mostgeneralapp}, with the difference that deviations
	during the first phase of play are punished by the \( T \)-pure minmax of
	the deviating player.
	Let \( \delta = \min_{i \in A_1} (u_i(\sigma_{n_0}) - n_0 \tilde{v}_i)
	\).
	The final phase is \( \left\lceil \frac{\card{S}+1}{\delta} \right\rceil
	\) independent repetitions of \( \sigma_{n_0} \) to punish any deviation
	by players in \( A_1 \).
	The proof that it is a \nashe{} follows from the fact that the \( T
	\)-pure minmax is both the punishment and the lower bound on the profile
	\( p \), and all players who are not best responding at all rounds are
	punished by \( \sigma_{n_0} \).
	Moreover, the equilibrium uses \( O(1) \) entropy for all players in \( T
	\) since all punishments in the first phase use strategy profiles where
	they do not mix, and the second phase is of length \( O(1) \).

	For the necessary condition, the result of Lemma~\ref{lmm:avgsubset}
	applied in the proof of Lemma~\ref{lmm:necfeasible} provides such a
	payoff profile.
\end{proof}
\end{document}